分类号：TP183　　　　　　　　　　　　　　单位代码：11232
　　　　　　　　　　　　　　　　　　　　　密　级：

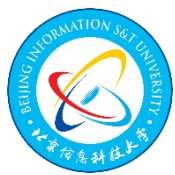

# 北京信息科技大学

专业硕士学位论文

# 铌酸锂易失性忆阻器的储备池计算

学　院：仪器科学与光电工程学院

专　业：电子信息

学　号：

作　者：赵元溪

学校指导教师：段文睿

企业指导教师：

完成日期：二零二三年四月二十日

# 摘 要

在传统数字计算机中，数据和信息以二进制数制的形式表示，并以静态的方式储存在晶体管的稳态中进行处理。然而，随着晶体管逐渐逼近其物理极限和冯·诺依曼瓶颈的到来，当前计算效率的提升速度也已经放缓。因此，从生物系统的动态和自适应特性中获得灵感进行神经形态计算研究具有很大的发展前景。本论文提出了一种新型铌酸锂（$LiNbO_3$，简称 LNO）忆阻器并应用于储备池计算系统，借鉴生物系统动态和自适应特性，通过该具有记忆功能的非线性电阻实现了储备池计算，相比于传统数字计算而言具有更强的计算能力和更高的能量效率。本文首先介绍了传统数字计算机的计算方式和现有计算效率的局限，然后介绍了基于忆阻器的动态特性进行计算的研究背景和目的。其次，详细介绍了本文所采用的忆阻器的制备流程和方法，并通过一系列电学测试验证其易失性、重复性和稳定性。随后，详细描述该忆阻器的性质及其在储备池计算系统中的应用。通过单个器件实现传统储备池计算中需要大量互连节点才能实现的功能，并通过 5 个忆阻器成功地完成了机写的阿拉伯数字图像的模式识别任务。最后，探讨了应用 LNO 忆阻器的储备池计算系统的研究成果对未来神经形态计算领域的理论和实际意义，以及可能的发展方向。

**关键词**：$LiNbO_3$忆阻器；易失性；储备池计算；神经形态计算；模式识别

# ABSTRACT

In conventional digital computers, data and information are represented in binary form and encoded in the steady states of transistors. They are then processed in a quasi-static way. However, with transistors approaching their physical limits and the von Neumann bottleneck, the rate of improvement in computing efficiency has slowed down. Therefore, drawing inspiration from the dynamic and adaptive properties of biological systems, research in neural morphology computing has great potential for development. Memristors, a class of nonlinear resistors with memory function, naturally embody dynamics through their internal electrical processes. This enables nonconventional computing paradigms with enhanced capability and energy efficiency, such as reservoir computing. In this paper, we propose a volatile memristor made of $LiNbO_3$ with nonlinear I-V characteristics and short-term memory function. This memristor is well-suited to be used as a nonlinear node in the storage layer of reservoir computing. With this system, we can achieve the same functionality as traditional reservoir computing with a single device, instead of a large number of interconnected nodes. The collective states of memristors after the application of trains of pulses to the respective memristors are unique for each combination of pulse patterns. This provides a more reliable and efficient way for subsequent output layer classification processing. The system was successfully used to recognize images of Arabic numerals 0-9. This work not only broadens the application scope of materials such as lithium niobate in neural morphology computing but also provides new ideas for developing more efficient neural morphology devices and systems.

# 目 录

# 第 1 章 绪论

## 1.1 研究背景及意义

计算机的运行和数据处理是基于二进制数制的，通过对二进制信号进行组合与处理能够实现数据存储、各种运算和逻辑功能。晶体管的稳态状态对应着二进制数值，其中导通表示为 1，截止表示为 0。通过控制晶体管的稳态状态，实现了各种计算和数据处理。因此，晶体管的发明和应用为计算机科学和信息技术的发展奠定了基础。

随着技术的发展，晶体管尺寸不断缩小，单个芯片上晶体管数量不断增加，计算机处理器变得更加高效和强大，计算能力得到了指数级的增长，这推动了数字计算在医学、生物学、国防等领域的发展，并产生了深远的影响[1, 2]。然而，在 2006 年之后，晶体管密度和计算效率的提升开始趋于饱和[3]，如图 1.1 所示。这种现象主要由两个原因引起。一方面，现代计算机采用冯·诺依曼架构，信息的存储和处理是分开的，这导致 CPU 等待内存存取的时间过长，影响了整个计算机系统的性能提升；另一方面，晶体管尺寸的缩小已经接近原子级别，进一步缩小尺寸会使得器件的性能不稳定，而芯片工艺技术的缓慢发展也制约着晶体管尺寸缩小的速度。因此，进一步提高晶体管密度和计算效率面临着重大的挑战。

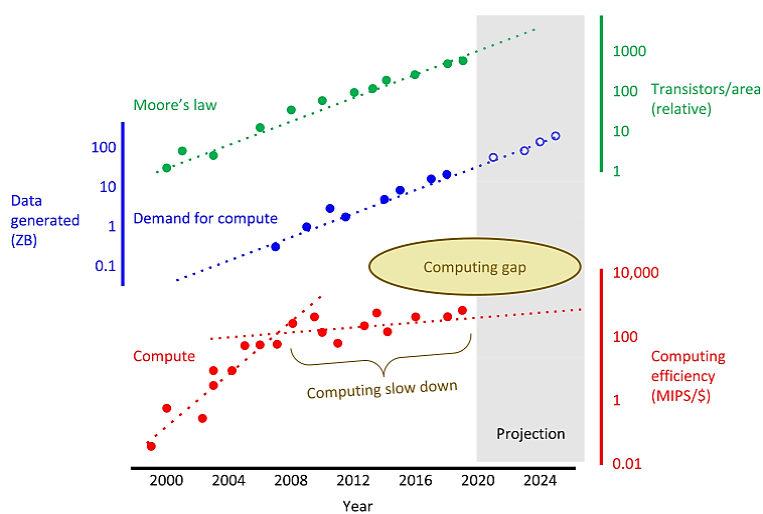

图1.1 近 20 年的晶体管密度、计算需求和计算效率变化趋势[3]

为了提高计算机系统的性能，有人提出了借鉴生物系统动态和自适应性的想法。生物系统在适应环境变化时，能够展现出相应的行为反应[4]。然而，计算机系统的硬件和软件都是静态元素。这种静态架构能够提供稳定的性能，但是缺乏动态和适

应性。计算机系统的动态和自适应性需要通过人工编程来调整指令，继而实现对不同需求的响应。为了解决这个问题，可以使用人工神经网络（Artificial Neural Network，简称 ANN）进行深度学习[5]。该方法利用训练样本和训练算法，动态更新网络权重，实现系统的自适应性。这种方法能够处理高维度和复杂的数据、非线性关系和分布式计算等，因此在许多应用领域表现出很优异的性能。

但值得注意的是，无论是在训练学习阶段还是在处理和应用阶段，计算机系统的基础硬件仍保持着静态、低复杂度的状态。因此，接下来的研究方向将从硬件层面开始，为人工神经网络提供自适应和复杂动态的能力。该研究方向旨在提高计算并行性、可扩展性、能源效率以及鲁棒性，从而实现进一步提高计算机系统性能和适应性的目标。

目前大多数模仿生物系统动态和自适应性的研究都是基于晶体管电路进行的。例如，IBM 的 Blue Gene/P 超级计算机拥有 147,456 个 CPU 和 144TB 的主内存，用于模拟猫的大脑皮层，但该计算机的功耗高达 2.9 兆瓦，存在功耗高和效率低的问题[6]。这种低效率现象是由晶体管静态的工作方式所造成的。晶体管在工作时只有开、关两种状态，其无法动态地适应环境变化。与之相比，生物系统具有强大的自适应和学习能力，能够灵活地适应环境变化。因此，研究人员开始探索使用基于新型器件的计算机系统来模仿生物系统的动态和自适应性。如果存在一种器件，能够在其内部展现出动态过程，那么每个器件就有可能在功能上替代复杂的数字电路[7, 8]。这种器件可以使用新型材料、新型结构或新型工作方式来实现，例如自旋电子器件、超导器件、基于光电效应的器件等。这些新型器件有望提供更高的计算并行性、更好的能源效率和更强的鲁棒性，从而提高计算机系统的性能、效率和可靠性。

忆阻器的诞生使得研究人员对上述新型器件的设想成为现实。忆阻器是一种带有记忆功能的自适应半导体器件，其内部物理过程是动态变化的。1971 年，蔡少棠教授依据对称性理论首次提出忆阻器的存在[9]，如图 1.2 所示。直到 2008 年，惠普实验室才首次制备出忆阻器[10]，如图 1.2 所示。

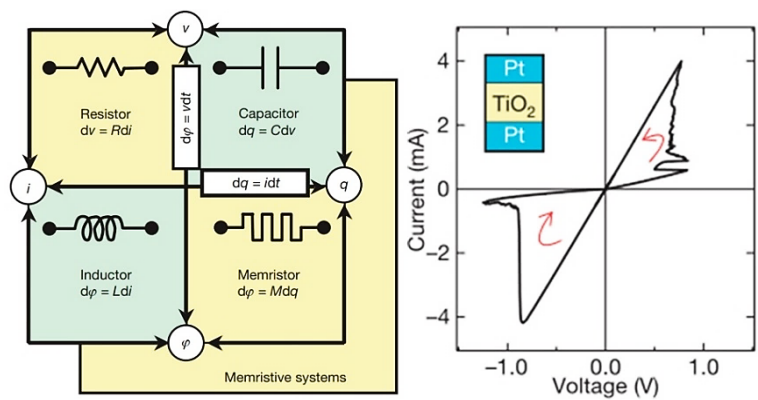

图1.2 忆阻器的模型提出和实现[10]

忆阻器的阻值由过去流经器件的电荷总量 $q$ 所决定，即由流经器件电流对过去时间的积分所决定。忆阻器是一种非线性电阻，具有记忆功能，可以自然地模拟生物系统的动态过程。由于其工作效率比 CPU 和 GPU 更高[11-17]，人们开始大量研究忆阻器，并发现了诸多有趣的应用，如模拟神经元和突触[11-14]、存储器[15]以及逻辑门[16, 17]等。

综上所述，利用具有动态特性的忆阻器可以取代由数百或数千个晶体管组成的复杂电路，同时合理运用忆阻器的物理特性和器件之间的相互作用，可以更加生动地模仿生物系统的动态和自适应性。

## 1.2 忆阻器

### 1.2.1 忆阻器的阻变特性

忆阻器是一种非常优异的自适应半导体器件，它具有记忆功能和动态特性，可以模拟生物系统的行为。与传统的线性电阻相比，忆阻器不仅可以表示当下电流与电压的关系，也能够包含过去的电流或电压的记忆，因此它是一个非线性电阻器件。忆阻器中的忆阻值 $M$ 是由公式（1.1）表示：

$$M(q) = \frac{d\Phi B}{dq} \tag{1.1}$$

根据法拉第电磁感应定律及复合函数求导法则，忆阻值可由公式（1.2）表达：

$$M(q(t)) = \frac{V(t)}{I(t)} \tag{1.2}$$

忆阻值会随着过去流经器件的电荷总量 $q$ 的累积而改变，因此忆阻器是一个非线性电阻器件，其伏安特性测试结果也表现出非线性曲线。忆阻器的初始状态为高电阻状态（高阻态，简称 HRS），但是对其施加电压时，它会转变为低电阻状态（低阻态，简称 LRS）。根据该器件在去除电压后能否保持 LRS，可以将忆阻器分为易失性和非易失性忆阻器。

图 1.3 中所示为理想形态的器件伏安特性曲线，注意该蝴蝶状的伏安特性曲线仅用作示意，并非真实测量结果。对于非易失性忆阻器而言，即使去除电压，仍可以保持在 LRS 状态。当再次施加与之前同向的电压时，器件的电阻将比初始状态更低。然而，要将器件从 LRS 转换到 HRS，则需要施加反向的电压。相应地，双向电压扫描的伏安特性曲线测量结果能够反应出该电学行为。与非易失性忆阻器相比，易失性忆阻器在去除电压后会在毫秒或纳秒内自动从 LRS 返回 HRS。因此，当再次施加

与之前相同方向的电压时，两次的电流响应曲线会基本重合。而对易失性忆阻器施加反向电压时，该器件的电阻变化与施加正向电压的情况相同，即从 HRS 转变为 LRS。

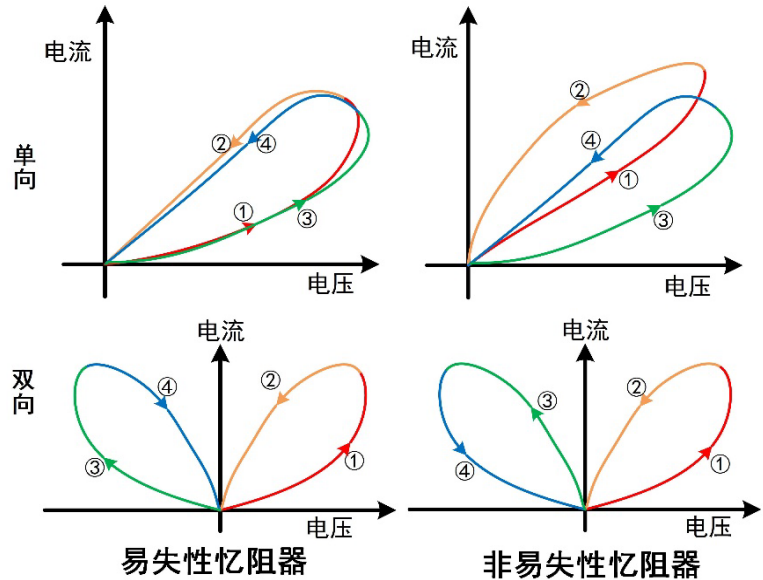

图1.3 易失性和非易失性忆阻器的理想伏安特性曲线

易失性忆阻器具有独特的短期记忆特点，这使得它在许多应用领域中表现出非常卓越的性能。例如，在计算机存储器领域，使用易失性忆阻器可以大大提高存储器的效率并降低功耗。在人工智能领域，易失性忆阻器可以用于构建神经形态系统，从而提高计算速度。

### 1.2.2 易失性忆阻器的阻变机理研究

易失性忆阻器的电阻变化可以由其内部的不同物理过程决定，其中包括离子效应、热效应和电子效应。然而，绝大多数易失性忆阻器的电阻变化是由离子效应主导的，其基本机理是在电场作用下，离子会发生定向移动，并在电极和忆阻层的界面处被氧化还原，最终形成导电细丝，使得器件的电阻值变低。当导电细丝断裂时，器件重新回到高阻态。这一过程可以通过 2014 年 H. Sun 等人制备的 $Ag/SiO_2/Pt$ 器件的扫描电镜图像中清晰地观察到[18]，如图 1.4 所示。

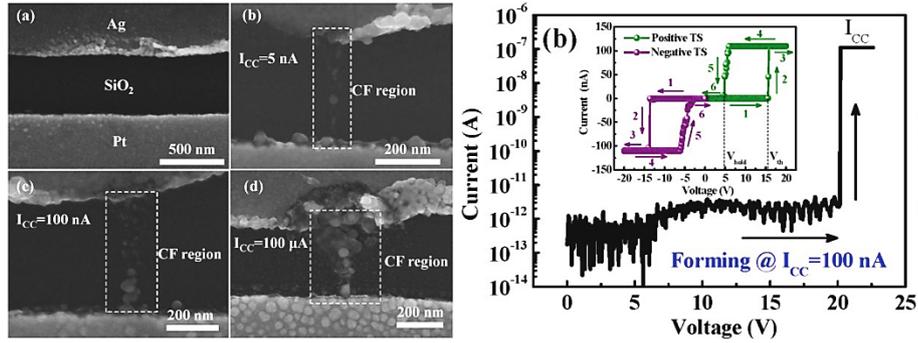

图1.4 Ag/SiO$_2$/Pt 器件的扫描电镜图像和伏安特性曲线[18]

除离子效应外，热效应也可导致忆阻器电阻的动态变化。在电场作用下，通过器件的电流会产生焦耳热，当积累到一定值时，热效应会导致忆阻材料的结构发生改变，从而引起器件电阻的变化。Mott 材料是一种存在上述现象的忆阻材料，例如 NbO$_2$ 和 VO$_2$。通过 M. D. Pickett 等人报道的一种 Mott 忆阻器存在这种机理[19]，其中 Mott 材料 NbO$_2$ 夹在两个 Pt 电极之间，如图 1.5 所示。当通过器件的电流足够大时，忆阻材料的温度超过 Mott 材料的转变值，这会导致材料由绝缘体向金属相变，进而在器件内部产生导电细丝。去除电压后，忆阻材料又恢复到绝缘体，从而使器件回到高电阻状态。

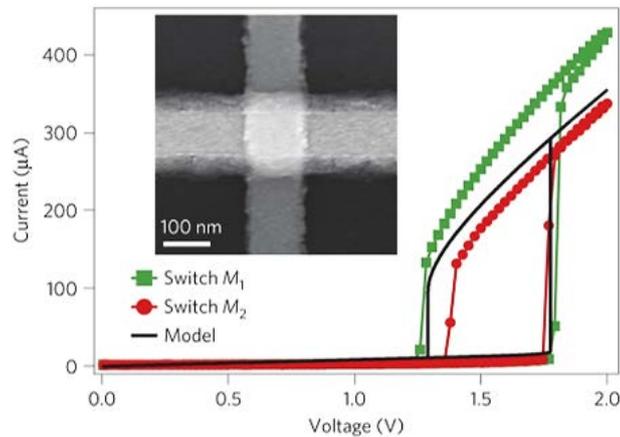

图1.5 Ag/NbO$_2$/Pt 器件的扫描电镜图像和伏安特性曲线[19]

基于电子效应的忆阻器电阻变化是由载流子填充忆阻材料中所有带电陷阱而导致的。Y. Kim 等人制备的 Pt/HfO$_2$/TiN 器件阐述了电子效应的阻变机理[20]，如图 1.6 所示。当施加电压时，带电陷阱能级会被拉到费米能级以下，从而导致电阻变小。而在去除电压后，载流子能够轻易地释放，从而使器件回到高阻态。

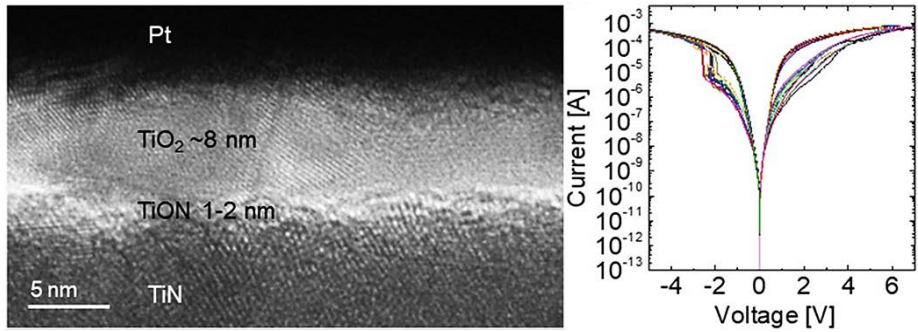

图1.6 Pt/HfO$_2$/TiN 器件的扫描电镜图像和伏安特性曲线[20]

易失性忆阻器的短期记忆和非线性特性使得它们具备在神经形态计算中发挥作用的潜力。因此，利用该器件动态特性的优势来构建具有动态和自适应特性的神经形态系统，可以克服传统数字计算的局限性。

## 1.3 易失性忆阻器的应用

### 1.3.1 神经形态的应用

人类的大脑是一个功耗极低且具有动态和自适应特性的器官，它包含着大约 $10^{11}$ 个神经元，每个神经元平均与 7000 个突触相连接。这些神经元是人类思维和行为的基本构建模块，每个神经元都可以与其他神经元通过突触连接起来，形成一个错综复杂的神经网络。利用易失性忆阻器可以模拟神经元和突触的行为，如图 1.7 所示。

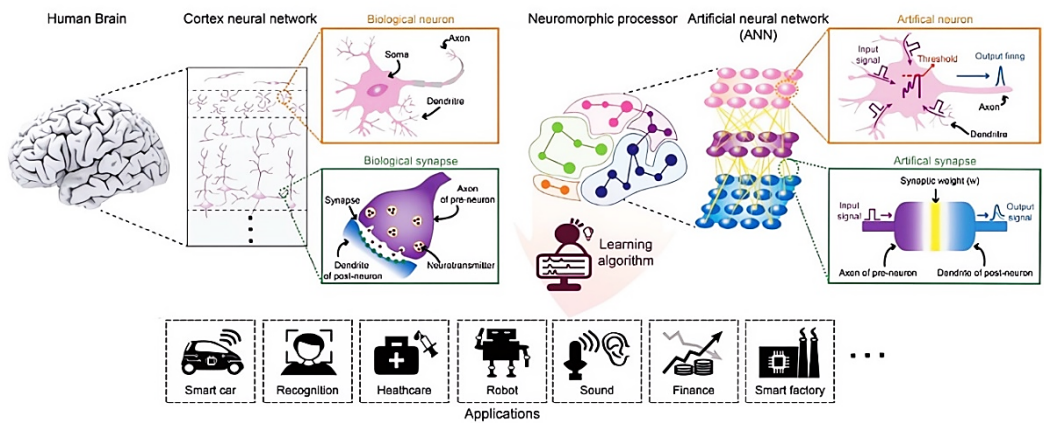

图1.7 忆阻器模拟神经元和突触行为[11]

神经元通过动作电位来传递信息，这种电位变化是由 Na$^+$和 K$^+$在内部流动所引起的。易失性忆阻器内部也有类似的离子动态迁移过程，因此易失性忆阻器可以用

来模拟人工神经元。W. Yi 等人通过制备 Pt/VO$_2$/Pt 的器件可以模拟 23 种类型的生物神经元动力学模型[21]，例如动作电位产生、尖峰时间依赖性、积分器和振荡器等。如图 1.8 所示，该图详细阐述了器件模拟神经元动作电位生成的基本过程。图中①过程，神经元保持静息电位，即 Na$^+$通道和 K$^+$通道都关闭。器件处于初始高阻态，漏电流流经器件产生 0.2-0.3V 的静息电位。图中②过程，Na$^+$通道激活引起超极化效应，使膜电位向负方向移动。图中③过程，K$^+$通道激活引起去极化效应，使膜电位向正方向移动。图中④过程，神经元逐渐恢复静息电位。此实验证明了由忆阻器、电容和电阻组成的人工神经元，可以模拟神经元动力学。

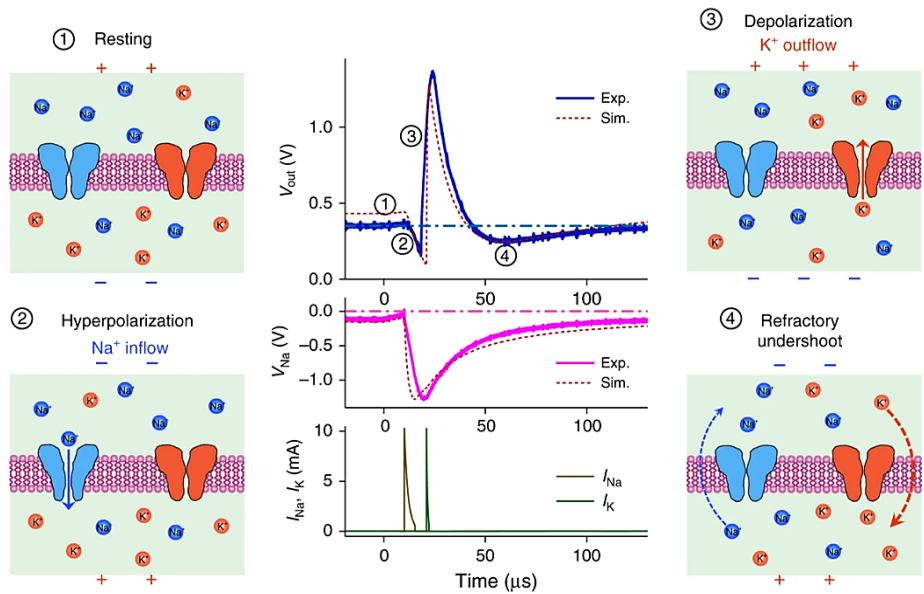

图1.8 Pt/VO$_2$/Pt 的器件模拟神经元动作电位产生过程[21]

在神经系统中，神经元是通过突触相互连接的，每个突触的功能都由其权重所衡量，该权重取决于神经刺激信号所释放的神经递质囊泡的数量和大小。易失性忆阻器的电阻变化，可以模拟突触权重的增加或减少，从而实现突触可塑性学习。S. La Barbera 等人制备 Ag/Ag$_2$S/Pt 忆阻器来模仿生物突触的突触可塑性学习[22]。该器件内部导电细丝的生长和断裂可以由施加的脉冲数量决定，从而控制器件的电阻变化。在长时程可塑性学习中，器件的电导变化如图 1.9 的 1、2 所示；在短时程可塑性学习中，器件的电导变化如图 1.9 的 3、4 所示。此实验证明了可以利用易失性忆阻器的电阻动态变化过程进行突触可塑性学习。

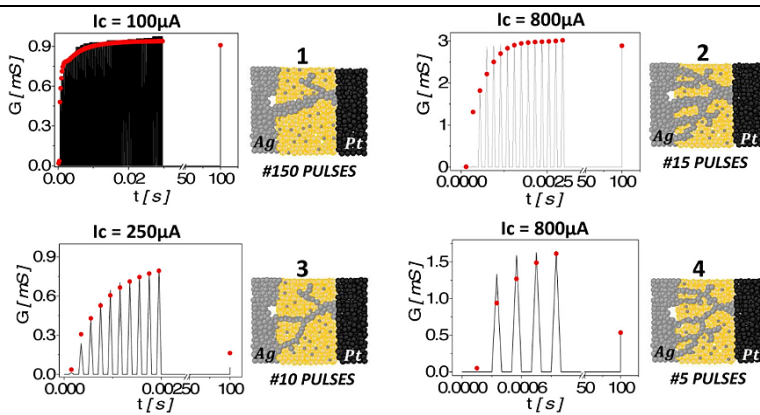

图1.9 Ag/Ag$_2$S/Pt 的器件模拟突触可塑性学习过程[22]

易失性忆阻器也可以利用其动态特性和非线性特性，应用在神经形态计算网络中。储备池计算（Reservoir computing，简称 RC）的概念是由 D. Verstraeten 在 2007 年提出的[23]，用来解决递归神经网络（Recursive Neural Network，简称 RNN）中出现的梯度消失和梯度爆炸等问题[24]。它是一种基于人工神经网络的计算框架，由三个部分组成，输入层、储层和输出层，其结构如图 1.10 所示。通过随机相互连接的节点组成储层状态，将输入的时间序列数据映射到高维空间中。储层中这些相互连接节点的权重是不变的，只需要训练输出层。因此与传统的 RNN 相比，RC 大大降低了训练成本。

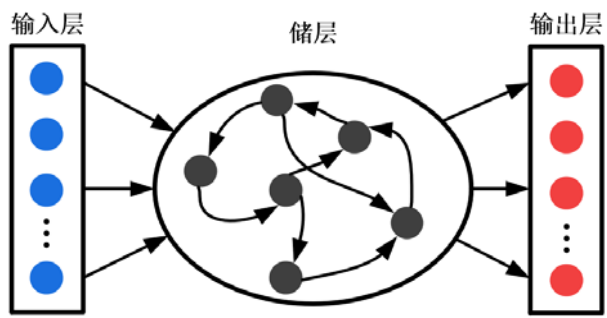

图1.10 储备池计算的基本结构

RC 具有两个关键特征：短期记忆和非线性特性。这些特征可以通过硬件器件来实现，例如忆阻器[25-31]、自旋电子振荡器[32, 33]、纳米线网络[34-36]和光电子器件[37-43]等，如图 1.11 所示。易失性忆阻器天然地具有这两个关键特征。器件的短期记忆使得器件的输出不仅由当前输入电压决定，而且还与前一刻的状态有关。因此，RC 可以获得丰富的储层状态。易失性忆阻器的非线性特性可以将输入空间的数据点映射到高维特征空间，使得在该空间中的输出是线性可分的。这使得 RC 更有利于后续的信息分类处理。此外，易失性忆阻器具有功耗低、响应快、易于集成等优点，因此非常适合应用于 RC 系统。

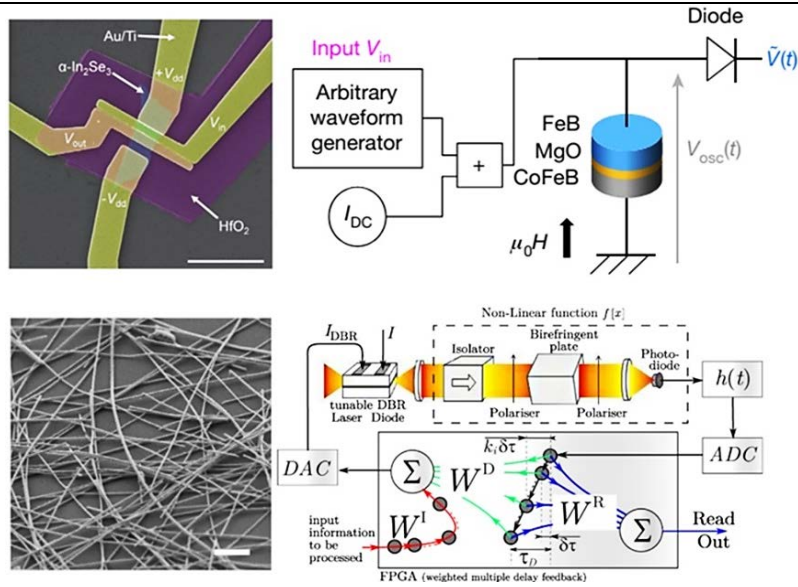

图1.11 可应用在 RC 系统的多类型硬件器件[27, 32, 36, 42]

J. Moon 等人利用 32×32 WO$_x$ 易失性忆阻器阵列构建 RC 系统[25]，如图 1.12 所示。利用器件内部的离子动力学和虚拟节点概念，成功实现了语音数字识别任务和混沌时间序列预测。该研究证明了通过储层映射输入的时间特征，即使是部分输入也能获得良好的分类效果。此外，通过定期更新储层以使其恢复到原始状态，即使在混乱的任务中，该系统也可以长期保持预测能力，无需重新训练。

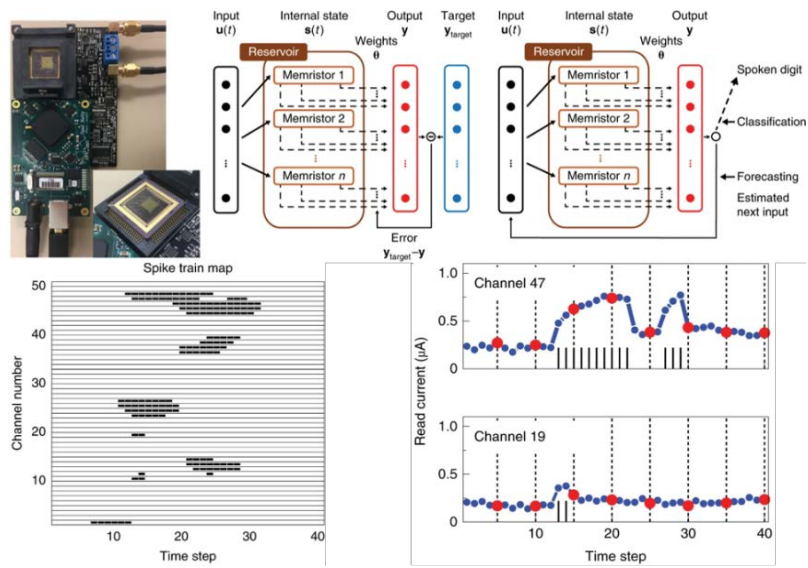

图1.12 基于 32×32 WO$_x$ 易失性忆阻器阵列的 RC 系统[25]

C. Du 等人利用 88 个 WO$_x$ 忆阻器构建 RC 系统[26]。该系统可以成功地完成手写数字识别任务，还能够解决二阶非线性动态问题，如图 1.13 所示。而且，即使传递函数形式是未知的，该系统也能够成功地得到预期的动态输出。

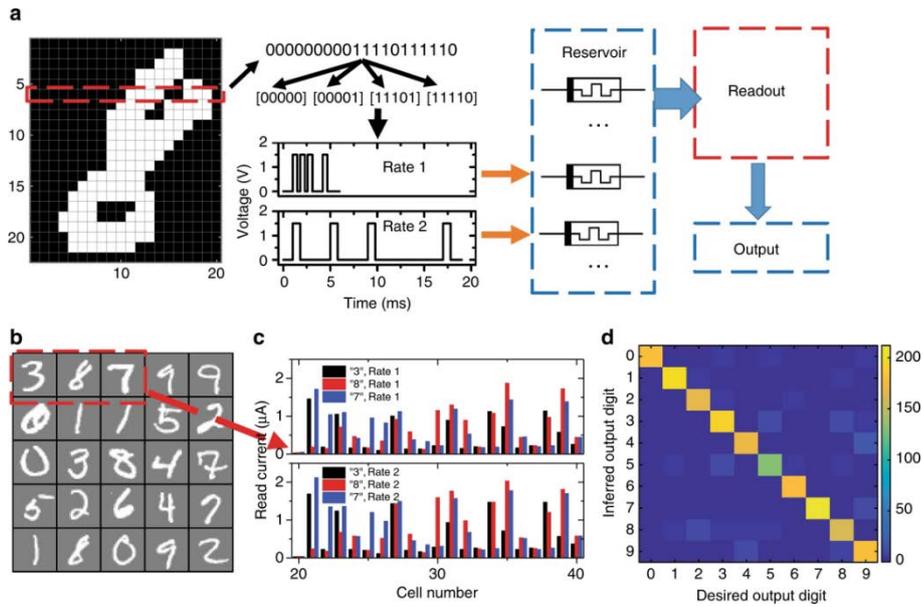

图1.13 基于 88 个 WO$_x$ 忆阻器的 RC 系统[26]

Keqin Liu 等人基于 α-In$_2$Se$_3$ FeSFET 构建 RC 系统[27]，如图 1.14 所示。该系统通过电阻匹配和电压分压，将 FeSFET 与在同一 α-In$_2$Se$_3$ 晶片上制造的平面器件堆叠成两个 RC。因为 RC 的输入和输出均为电压，所以可以级联使用。与单层 RC 相比，多层 RC 系统表现出更大的存储量。该系统通过实现时间序列预测和波形分类任务，证明了多层架构的存储量和计算能力。

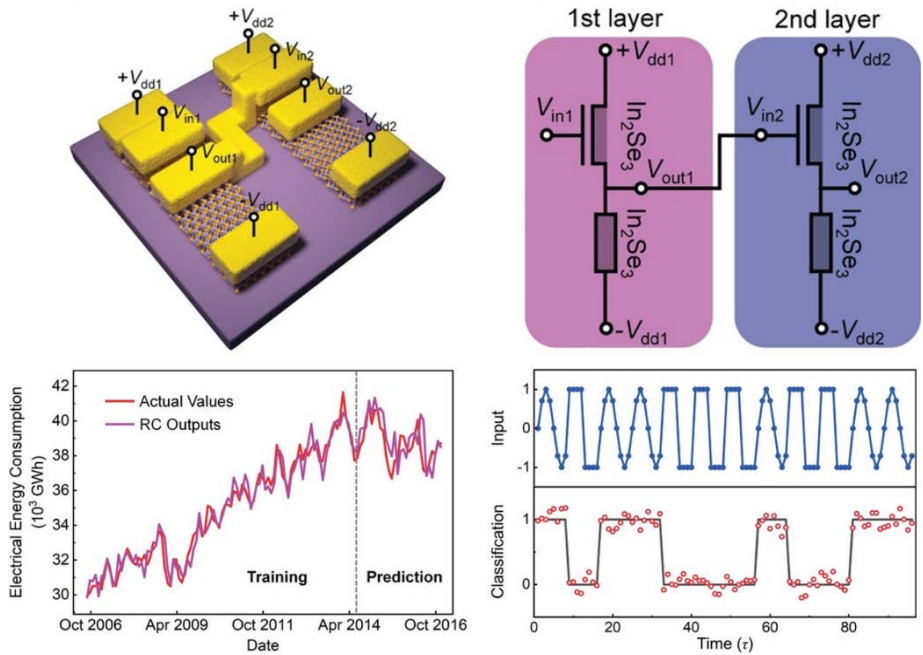

图1.14 基于 α-In$_2$Se$_3$ FeSFET 的 RC 系统[27]

Yanan Zhong 等人制备的 RC 计算系统包含两种类型的忆阻器，如图 1.15 所示。其中 Ti/TiO$_x$/Pd 易失性忆阻器用作非线性节点，而 TiN/TaO$_x$/HfAlO$_y$/TiN 非易失性忆

阻器则用于输出层[28]。该系统可用于实时处理时空信号，其功耗比传统 RC 系统低三个数量级以上。不仅如此，该系统还成功完成了心律失常检测和动态手势识别任务，其结果的准确率分别高达 96.6％和 97.9％。

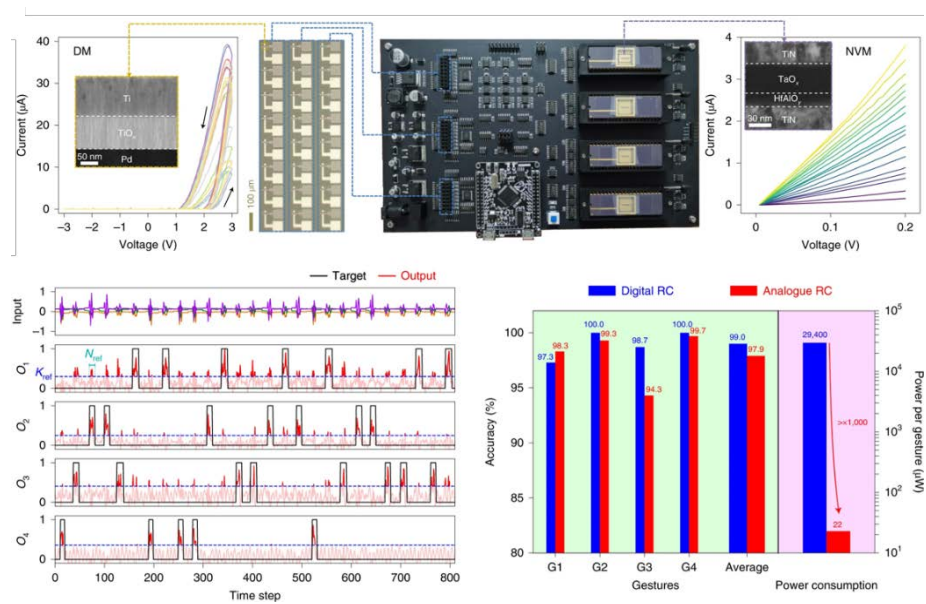

图1.15 基于 Ti/TiO$_x$/Pd 易失性忆阻器和 TiN/TaO$_x$/HfAlO$_y$/TiN 非易失性忆阻器的 RC 系统[28]

综合上述研究可以发现，基于易失性忆阻器构建的储备池计算系统，利用了器件的短期记忆和非线性特性，这可以使该系统具有更大的存储量、更强的计算能力和更低的功耗。同时，该系统在语音数字识别、波形分类、混沌时间序列预测和神经信号分析等领域表现优异。

### 1.3.2 选通器件的应用

在存储器阵列中，将信息写入某个存储器时，由于阵列结构的原因，可能会引起一些不希望被写入的存储单元发生写入现象，这种"潜行"到原本不该写的存储单元产生的电流被称为潜行电流。虽然潜行电流可以实现信息存储和神经形态计算应用中的随机性，但是也会带来读取精度降低、数据丢失和功耗高的问题。目前，抑制潜行电流并实现忆阻器阵列的可行办法之一是将忆阻器和晶体管组合构建 1T1R 结构[44]，然而这两者的结合不仅增加了面积，还会阻碍忆阻器发挥出尺寸小的优势。另一种方法是利用易失性忆阻器作为选择器，将其作为选通器件与存储单元串联。该方法能够使其具有优良的选择性和低漏电流，可以降低潜行电流和功耗[45-48]。

Qilin Hua 等人制备 Ag nanodots/HfO$_2$ 作为选择器，TaO$_x$/Ta$_2$O$_{5-y}$ 作为阻变存储器，将两者结合构成 1S1R 结构[49]，如图 1.16 所示。该选择器具有 10$^9$ 开关比，低于 1pA

的漏电流以及高达 200°C的良好热稳定性。1S1R 器件也证明了其具有抑制潜行电流的优异表现和 $10^8$ 高循环稳定性。

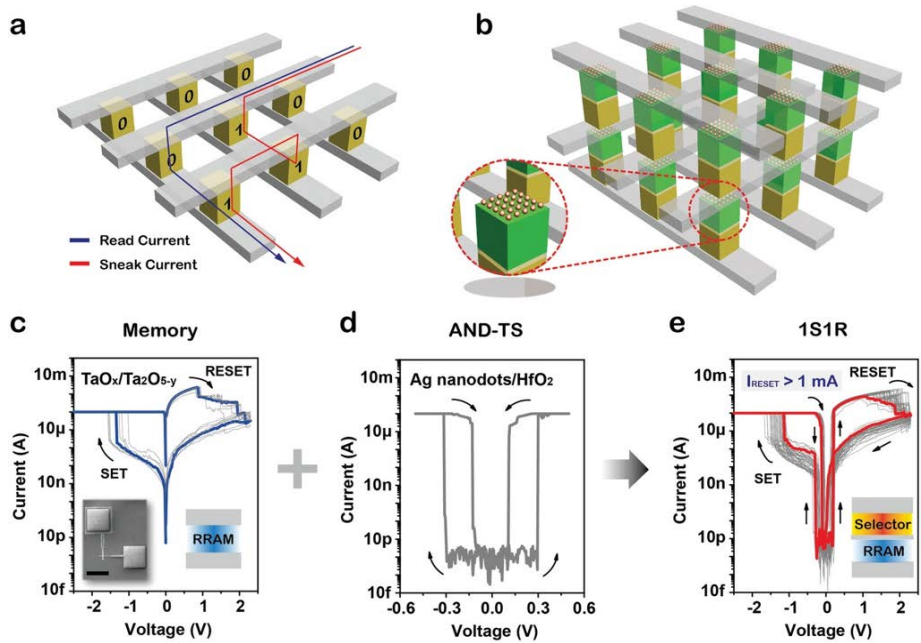

图1.16 Ag nanodots/HfO$_2$ 选通器件及其伏安特性曲线[49]

## 1.3.3 硬件安全模块的应用

由于日常生活中信息爆炸式增长，信息安全已成为当前热门话题。在保护通信安全方面，随机数生成器（RNG）起着至关重要的作用。RNG 可以分为伪随机数生成器（PRNG）[50]和真随机数生成器（TRNG）两种类型。与 PRNG 相比，TRNG 具有更高的熵质量和更好的安全性能。易失性忆阻器因其内部具有独特的物理过程可以作为随机性的来源，因此被广泛用于实现 TRNG。

Hao Jiang 等人研制了一种基于 Au/Ag/Ag:SiO$_2$/Pt 易失性忆阻器的 TRNG[51]，如图 1.17 所示。该器件利用忆阻材料中金属原子的扩散动力学来产生随机性，并通过比较器、与门和计数器等组件构建 TRNG。经过 15 项 NIST 随机性测试，该器件表现出良好的随机性，将其集成到存储系统中，可提高系统的安全性，同时降低功耗。

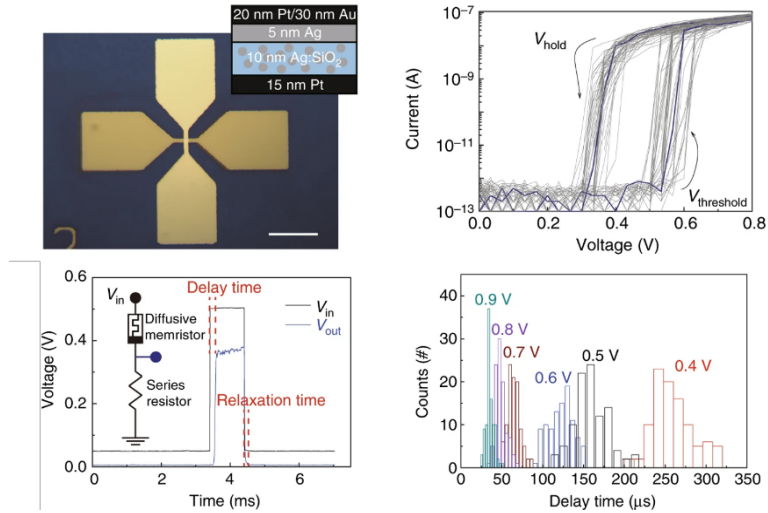

图1.17 Au/Ag/Ag:SiO$_2$/Pt 易失性忆阻器及其电学测试结果[51]

## 1.4 论文组织与架构

忆阻器是一种新型的具有广泛应用前景的自适应半导体器件，然而，忆阻器在实际应用中仍面临着一些重要挑战：1.制造工艺难度大：忆阻器的制造工艺较为复杂，需要精密的工艺控制和高端的设备，这对制造成本和生产效率都会带来一定的影响。2.稳定性不足：忆阻器的稳定性对其性能和寿命有很大影响。由于其特殊的物理机制和材料结构，忆阻器的稳定性相对较低，容易受到温度、压力等环境因素的影响，从而导致器件性能的下降和寿命的缩短。3.计算和控制难度高：忆阻器的物理机制比较复杂，需要通过复杂的计算和控制方法来实现其特定的功能。这对芯片设计和系统集成提出了更高的要求，也增加了应用的难度和成本。4.大规模应用受限：目前忆阻器的生产和应用还处于起步阶段，其规模和可靠性都需要进一步提高，才能在大规模的应用中发挥出其优势。此外，忆阻器在一些特定应用场景下，如高温、高辐射等极端环境下的应用，也面临着一定的技术难度和风险。

为应对和解决这些挑战，本文将用五章内容对LiNbO$_3$忆阻器（LNO忆阻器）的储备池计算系统的构建和性能分析进行阐述，论文组织框架如图1.18所示。

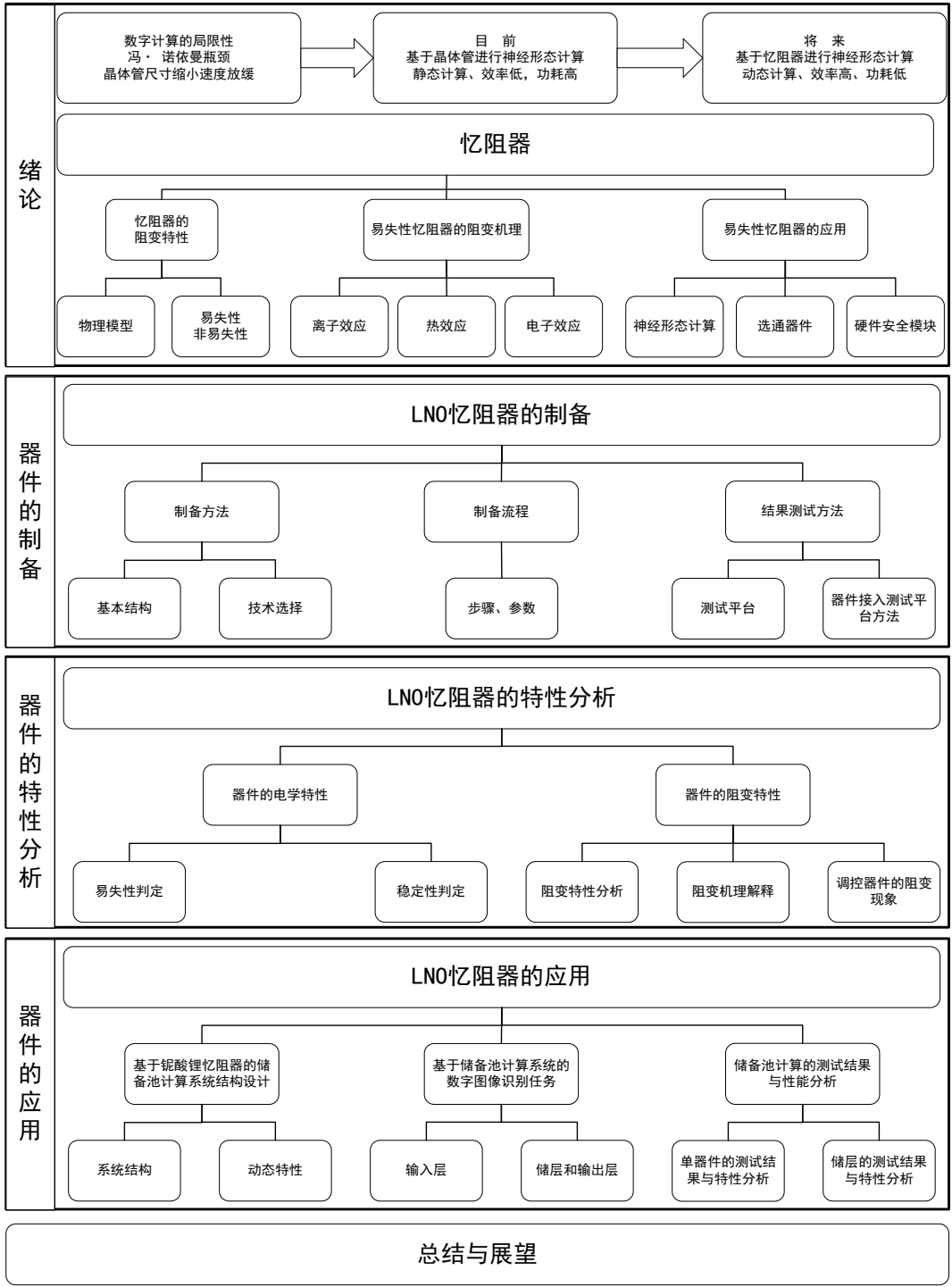

图1.18 论文框架

具体安排如下：

第 1 章简单地概括本课题的研究背景与意义。介绍忆阻器的来源、实现以及分类，着重阐述了易失性忆阻器的特点、阻变原理和应用，并对本文的结构进行安排。

第 2 章介绍 LNO 忆阻器的制备过程。首先提出器件的结构设计，其次进行制备仪器的选型，并详细描述了制备全过程，最后阐述电学测试方法及仪器。

第 3 章分析 LNO 忆阻器的阻变特性。首先进行电学性能测试，其次对测试结果进行分析，对器件的阻变原理进行深入分析。最后通过脉冲调控实现器件的电阻变化。该器件展现出良好的易失性和较强的电阻可控能力。

第 4 章利用 LNO 忆阻器进行神经形态计算应用的研究。利用器件构建储备池计算系统，通过完成 0-9 数字图像识别任务，对系统的功能进行验证。结果表明，易失性器件的短期记忆和非线性特性在储备池计算系统中发挥着重要作用。

第 5 章总结本论文的研究内容，并对易失性忆阻器的后续研究进行了展望。

## 1.5 本章小结

本章介绍了传统数字计算的发展受限于冯·诺依曼瓶颈和晶体管尺寸缩小速度放缓两方面的因素。为了克服这些问题，研究人员从生物系统的动态和自适应特性获得灵感，开展神经形态计算学习的研究。然而，传统的基于晶体管的研究存在着静态计算、效率低和功耗高等问题。相反，利用忆阻器进行研究则具有动态计算、效率高和功耗低的优势。本章还介绍了忆阻器的来源及其两种类型，即易失性忆阻器和非易失性忆阻器，并详细描述了易失性忆阻器的阻变原理和应用领域。基于这些内容，本文进一步展开了全文的规划，构建了全文的框架。

在电子信息科学中，忆阻器是一种重要的电子器件。然而，它面临着制造工艺难度大、稳定性不足、计算和控制难度高和难以规模应用等挑战。为了解决这些挑战，后面将逐章展开说明。

# 第 2 章 LNO 忆阻器的制备和测试方法

第 1 章介绍了忆阻器具有多种功能，接下来将详细介绍 LNO 忆阻器的器件结构、制造仪器、工艺过程和测量方法。具体来说，忆阻器的器件结构主要包括阻变材料和电极。阻变材料是忆阻器的关键部件，它的物理和化学性质直接影响着忆阻器的性能。在制造过程中，本文确定选取 LNO 材料，并通过一系列的工艺步骤来完成器件的制备。制造设备则是完成这些工艺步骤的关键工具，它们的精度和稳定性对于器件的性能影响巨大。同时，在制备过程中需要精确地控制工艺参数，以确保器件的稳定性和一致性。为了进一步了解忆阻器的性能和特点，需要进行各种表征测量。常见的表征测量方法包括电学特性测试、结构表征、显微成像等。其中，电学特性测试是评价忆阻器性能的主要手段，通过测试其电阻、导电性、记忆性等参数，可以判断器件的稳定性和性能优劣。

## 2.1 LNO 忆阻器的制备方法

### 2.1.1 器件基本结构

LNO 忆阻器是一种基于钙钛矿结构的忆阻器，具有优异的电学性能和忆阻效应。目前，忆阻器的结构主要包括三明治结构、三端结构和多层结构等，其中三明治结构由于其简单的结构，制备流程和工艺需求大大简化，降低了制造复杂度和成本，因而具有明显的优势。其制备工艺流程的简化，使得忆阻材料受制备电极的影响较小，忆阻材料质量好，忆阻器的性能也更有保障。此外，三明治结构还能避免耗散、尺寸效应等问题，提高器件的可靠性，减少出现退化、失效等情况的可能性。因此，本论文采用了制造过程简单、制造成本低、可靠性高的三明治结构作为 LNO 忆阻器的结构。

本论文制备的 LNO 忆阻器的整体结构如图 2.1 所示。在 Si/ SiO$_2$ 基底上制备底电极金（Au）、忆阻材料 LNO 以及顶电极铂（Pt），三种材料均采用了薄膜形态，膜厚尺度为十~百纳米级。这种结构具有制备工艺简单、容易实现、可靠性高等优点，能在尽量减少工艺流程和制造成本的同时，实现良好的忆阻效应和电性能。

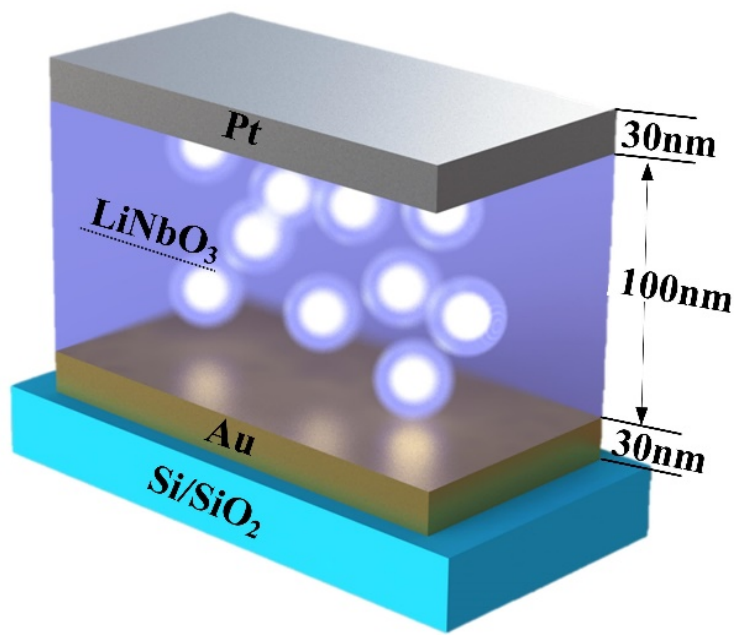

图2.1 LNO 忆阻器结构示意

## 2.1.2 制备技术选择

LNO 忆阻器的三种材料均为纳米级的薄膜材料，器件的性能指标与构成器件的材料质量有着密切的关系，因此选择合适的沉积薄膜的方式尤为重要。制备薄膜材料的方法有很多种，根据工作原理可以大概分为两类，PVD 物理气相沉积法和 CVD 化学气相沉积法。CVD 化学气相沉积的原理是在一定温度和反应气体条件下，利用等离子体激活、高温等方法，使反应物质发生化学变化，生成固态物质沉积在基片表面。这种方法适用于制备碱及碱土类以外的金属（Ag、Au 困难）、金属化合物和合金等。PVD 物理气相沉积的原理是在真空环境下，利用溅射、蒸发和电离等之类的物理气相方法，使反应物质在基片上凝聚，并沉积成薄膜。这种方法适用于制备除 C、Ta、W 所有固体、卤化物和热稳定化合物等。

LNO 忆阻器的三种薄膜材料分别为 Au、LNO 和 Pt。由于 CVD 化学气相沉积法无法用于制备 Au 和 LNO 碱类金属的薄膜材料，因此采用 PVD 物理气相沉积法作为研究方法。考虑到实验室条件和成本等因素，选择磁控溅射法来沉积薄膜材料。这种方法有许多优点，比如能够在较低的温度下制备高熔点材料的薄膜，具有较大的沉积面积，以及薄膜附着力强、组分稳定、膜层均匀致密等特点。因此，磁控溅射法是制备器件薄膜材料的最佳选择。

磁控溅射的工作原理是将欲沉积的材料制成靶材固定在阴极上，然后在靶面和基片之间施加几千伏电压，并通过电子与氩原子的碰撞产生溅射原子，最终在基片表面沉积成膜的过程。

使用 AJA 磁控溅射仪制备器件的三种薄膜材料，该仪器如图 2.2 所示。在沉积薄膜时，需要注意三个关键参数：工作功率、工作气压和沉积速率，它们会影响薄膜的质量和生长速率。这三个参数应该根据材料的性质进行选择，并可在仪器的软件控制界面中进行调节，如图 2.2 所示。

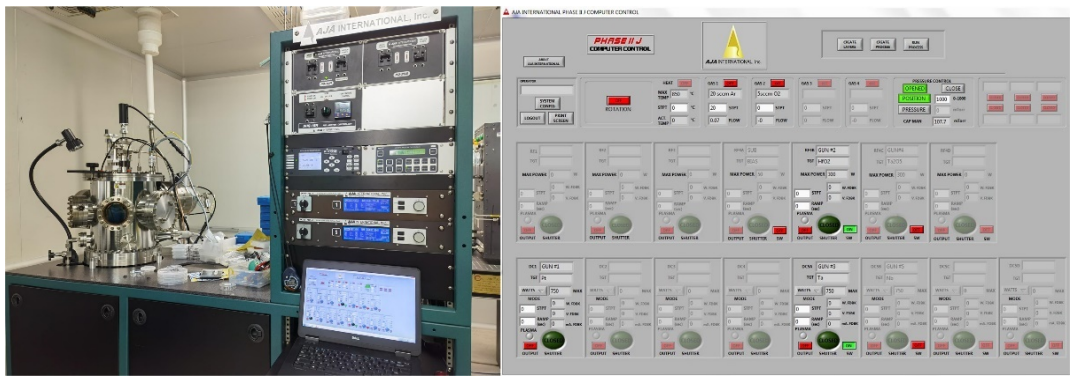

图2.2 磁控溅射仪及其典型参数设置

本论文采用曝光技术，将 LNO 忆阻器的结构设计和图形信息转移至基片上。微纳制造工艺中的曝光技术通常分为三类：光学曝光、电子束曝光和离子束曝光。其中，电子束曝光和离子束曝光是通过加速电子和离子并将它们聚焦成束来进行曝光的技术。虽然电子束曝光和离子束曝光具有非常高的分辨率和控制精度，可以制造高质量和复杂的芯片结构，但是其设备和操作成本较高，操作过程难以控制，曝光速度也较慢，只适用于小批量生产。相反，光学曝光的设备操作简单，可以制造大规模的器件结构，具有较低的成本和高效率。因为 LNO 忆阻器的三明治结构较简单且无需复杂的图形信息，所以本论文选择光学曝光技术将器件的图形信息转移到基片上。光学曝光是通过光学系统将掩模上的图案映射到光刻胶上，然后使用紫外线等光源对光刻胶进行曝光的技术。其原理示意图如图 2.3 所示：透过掩模的透明部分，光线得以穿过形成图案；而不透明部分则会阻挡光线的传递。通过选择合适的光学系统，将图案按比例缩小并转移到需要制作的芯片上，随后在光刻胶表面形成相应的芯片结构。

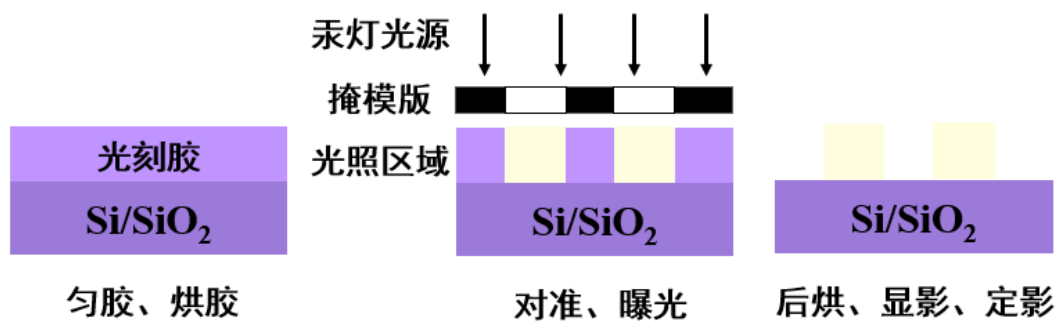

图2.3 光学曝光流程示意

使用光学曝光技术转移器件图形信息的流程为匀胶、烘胶、对准、曝光、后烘、显影和定影。涂覆光刻胶和烘胶的过程采用匀胶仪和烘胶台完成；光学曝光过程采用德国 SUSS mA8 双面光刻机实现。显影液和定影液被用于去除未暴露于光的区域。上述仪器的实物图如图 2.4 所示。该光刻机配置基于 Windows 操作系统的图形化控制软件，具有工艺编程、设备控制、硬件自我故障诊断等功能，操作方便快捷。曝光光源为汞灯光源，波长为 350-450nm，可支持恒定光强和恒定功率模式曝光。光刻机配置消衍射及微镜式光学系统，可以在同一设备上实现"高分辨率"和"大景深"两种模式曝光，并且可以随时切换。此外，曝光模式可支持硬接触、软接触、接近式和真空接触，分辨率优于 1.5μm。该光刻机可满足制备 LNO 忆阻器的低成本、高效率需求。

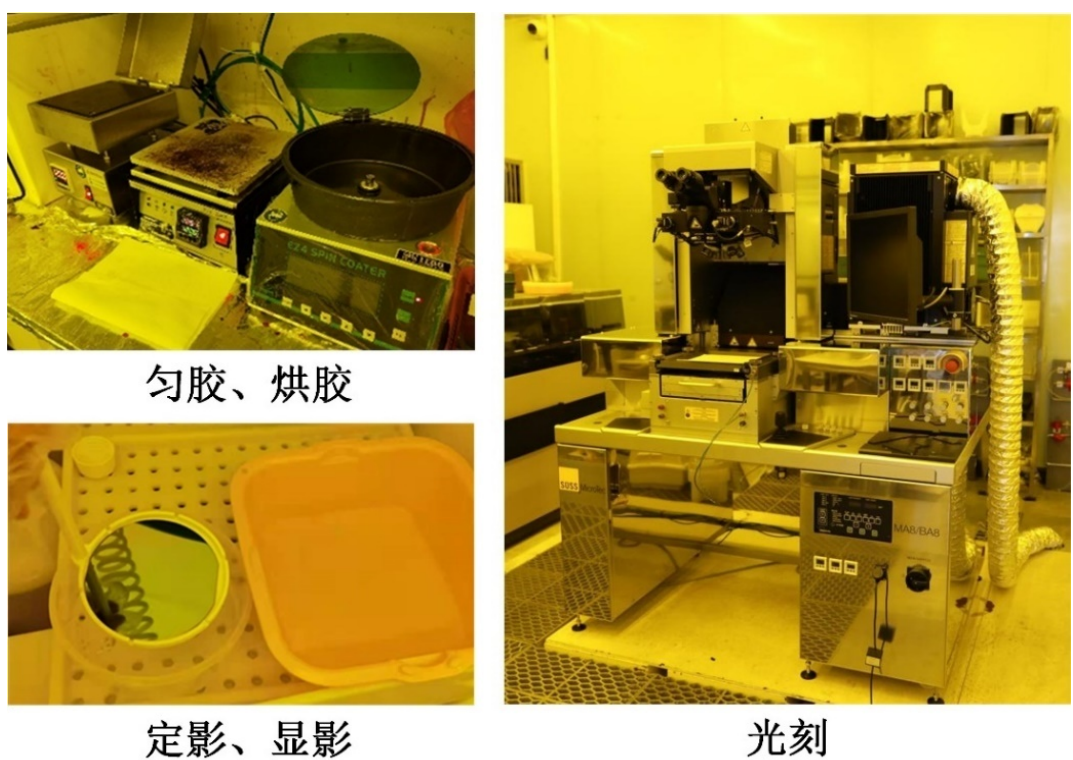

图2.4 光学曝光仪器

在制备 LNO 忆阻器的过程中，采用光学曝光技术，需要使用一个掩模版来承载器件的全部信息。器件的形状和结构设计以图形的形式制作在掩模版上，以确保准确传递所需的信息。对于三明治结构的 LNO 忆阻器件而言，其制备过程需要进行两次加工。首先需要制备底电极，因此需要使用一个包含底电极图形信息的掩模版，如图 2.5 中蓝色掩模版。接着，还需要进行第二次加工，即制备忆阻材料和顶电极，此时需要使用一个包含顶电极结构图形信息的掩模版，如图 2.5 中红色掩模版。

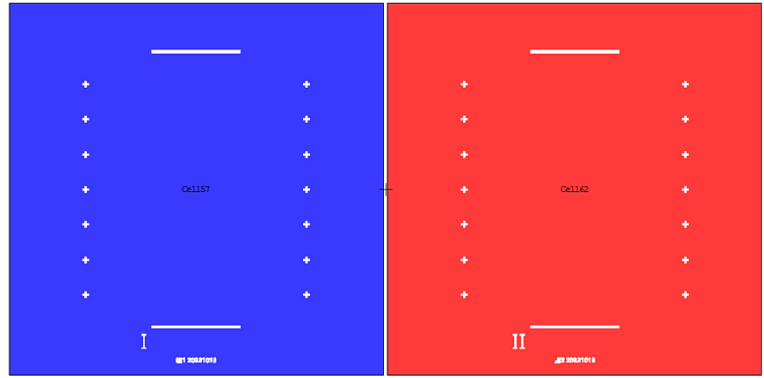

图2.5 掩模版设计图

本论文根据实际器件面积,选择了 5 英寸的掩模版。在制备器件时,选择掩膜对准式曝光的方式,因此掩模版图形和光刻图形的尺寸相同,底电极形状示意图如图 2.6 所示。

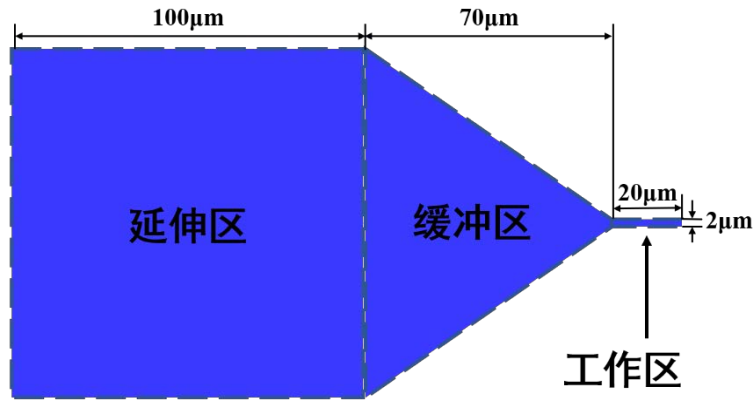

图2.6 底电极形状示意

两个电极形状设计相同但方向不同,图中电极的工作区域长度为 20μm,宽度为 2μm。电极延伸区域是一个长为 100μm 的正方形,连接工作区域和延伸区域的是缓冲区。缓冲区域是底为 100μm 高为 70μm 的三角形。按照上述尺寸规格,分别绘制底电极掩模版和顶电极掩模版,并将两个掩模版中的图形套刻后,可以得到十字交叉式忆阻器结构。

为了探究器件的阻变机理,需制备多个尺寸的器件,并检验其电学特性与器件面积的相关性。因此,绘制了多个宽度的正方形电极,宽度分别为2μm、4μm、8μm、16μm 和 32μm,如图 2.7 所示。

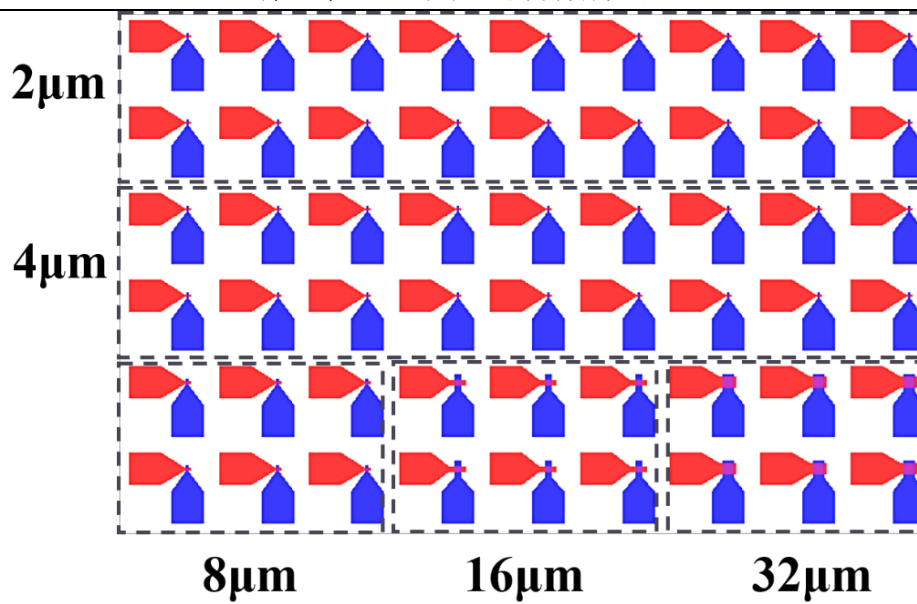

图2.7 多尺寸忆阻器套刻布局

在掩模版的两侧,有两排对准标记,这些对准标记主要用于第二次光刻的套刻过程。设计图如图 2.8 所示。

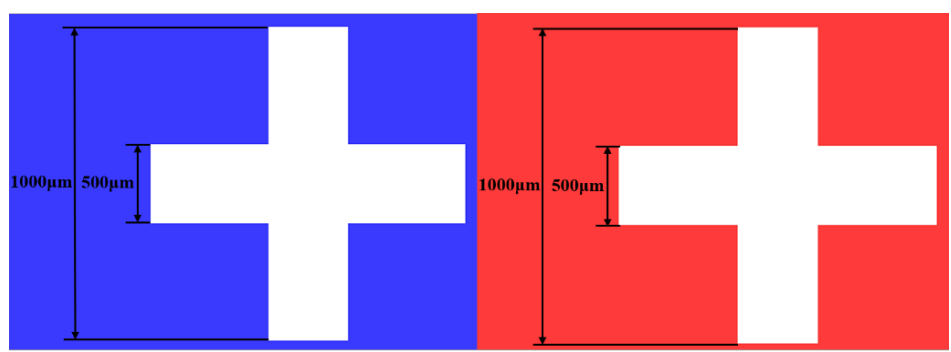

图2.8 对准标记

图案为两个宽为 500μm,长为 2000μm 的长方形垂直交叉组成的十字架,两个掩模版上的对准标记图案和位置均相同。在装载顶电极掩模版后,通过光刻机的对准系统,先对对准标记区域进行低倍数电镜放大观察,进行掩模版和基底硅片在 X 和 Y 方向上的粗对准。接着,通过高倍数电镜的放大观察,将两个十字对准标记重合,完成图形的套刻。

## 2.2 LNO 忆阻器的制备流程

制备 LNO 忆阻器的流程包括通过光刻和沉积薄膜的方法,在基底上制备底电极、LNO 薄膜和顶电极。首先,需要选择合适的基底材料,如硅片,并清洗和处理其表

面，使其平整、光滑且无杂质。接下来，利用光刻技术制备所需的底电极结构，并在基底表面生长 Au 薄膜，采用物理气相沉积方法，并控制反应条件以获得良好的电极结构和接触。然后，使用光刻技术将顶电极结构转移到已有 Au 薄膜的基片上，并在此基础上再生长 LNO 薄膜和 Pt 薄膜。制备该器件工艺流程如图 2.9 所示。

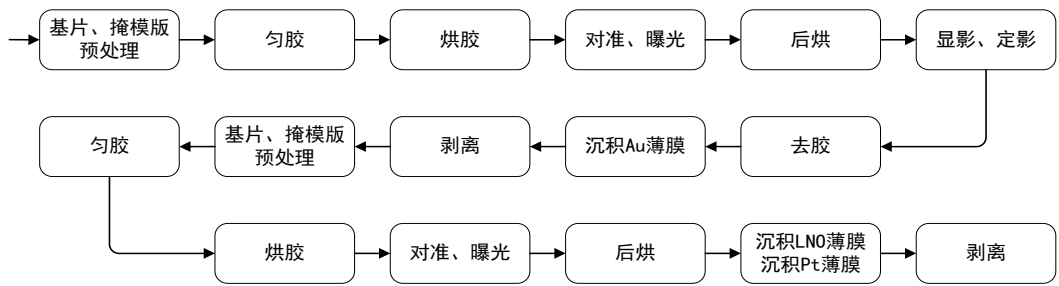

图2.9 制备 LNO 忆阻器的工艺流程

具体步骤如下：

（1）选取 4 寸硅片作为基片，对底电极掩模版和基片进行预处理。依次使用丙酮、酒精和去离子水清洗掩模版和基片。随后将它们放入 100°C 的烘箱中，烘干表面液体。这个过程的目的是去除表面杂质，例如颗粒、有机物、工艺残余和可动离子，以及水蒸气。

（2）使用匀胶仪来涂覆 AZ GXR-601 光刻胶，具体分为 3 个步骤：a）放置托盘和硅片并打开真空泵，使硅片吸附在托盘上方。b）在硅片表面滴入适量的光刻胶溶液。c）设置转速及时间进行涂胶，通常需要进行两次，第一次的匀胶速度为 700r/min，持续时间为 9 秒，第二次的匀胶速度为 3000r/min，持续时间为 20 秒。在此操作下，光刻胶会向着硅片外围移动，也可以称为甩胶。经过甩胶之后，留在硅片表面的光刻胶量不到原有量的 1%。

（3）使用烘胶台对基片进行前烘处理。前烘的目的是去除光刻胶中残余的溶剂，提高光刻胶的粘附性，释放光刻胶内部的应力，并防止光刻胶污染光刻机。将烘胶台的温度设定为 100°C，将基片放置烘胶台上烘烤 90s。

（4）使用德国 SUSS mA8 光刻机进行光刻工艺。打开光刻机的氮气、真空和压缩空气阀，汞灯起辉预热 30min。将底电极掩模版和基片固定在装版架。曝光参数设置为接近式曝光，曝光强度为 200Dose，持续时间为 20s 进行曝光。光刻胶中的感光剂在受到特定波长的光照射后发生光化学反应，从而使掩模版上的图形被复制到基片上。

（5）使用烘胶台对基片进行后烘处理，这一步可以提高后续显影速度。将基片放置在烘胶台上 100°C 烘烤 90s。

（6）为了去除正光刻胶的感光区域，进行显影和定影过程。将基片置于清洗架浸没于显影液（MF 319 DEVELOPER）中持续晃动 45s，再浸泡在定影液（去离子水）中，取出基片用氮气枪吹干表面液体。

（7）使用去胶机除去基片表面残余的光刻胶。将基片放置在等离子去胶机中，通 28sccm 氧气和 280sccm 氩气，功率为 100w，持续时间为 30s。

（8）使用 AJA 磁控溅射仪沉积底电极 Au 薄膜。将基片和 Au 靶材分别固定在阳极和阴极，腔体抽真空至真空度 $3\times10^{-7}$Torr，基片台开始旋转，通 20sccm 的氩气，设置 30mTorr 使靶材起辉。起辉成功后，设置工作气压为 3mTorr，工作功率为 30w，根据 Au 薄膜的溅射速率，沉积薄膜的时间为 3min，薄膜厚度为 30nm。完成沉积薄膜后，腔内通氮气直至腔内恢复至大气压下，取出基片。

（9）使用丙酮剥离未曝光区域沉积的薄膜。将基片浸泡在丙酮溶液中，剥离，用酒精和去离子水冲洗基片，氮气枪吹干表面液体。

（10）重复步骤（1-3），对已沉积 Au 薄膜的基片和顶电极掩模版进行预处理、在基片上匀胶和烘胶。

（11）使用光刻机进行光刻工艺。通过掩模版和基片上的十字对准标记进行套刻，将顶电极掩模版上的图案转移到基片上，曝光参数设置为接近式曝光，曝光强度为 200Dose，持续时间为 20s 进行曝光。

（12）重复步骤（5-7），去除未曝光区域的光刻胶。

（13）使用磁控溅射仪沉积 LNO 薄膜，工作气压为 3mTorr，工作功率 70W，溅射时间为 20min，薄膜厚度为 100nm。

（14）使用磁控溅射仪沉积顶电极 Pt 薄膜，工作气压为 3mTorr，工作功率 30W，溅射时间为 3min，薄膜厚度为 30nm。

（15）溅射后，将基片浸泡在丙酮溶液中，剥离未曝光区域沉积的薄膜，用酒精和去离子水冲洗，氮气枪吹干。至此 LNO 忆阻器制备完成。

## 2.3 LNO 忆阻器的制备结果测试方法

### 2.3.1 测试平台

本论文采用了 Cascade 探针台、Keysight B1500A 半导体参数分析仪和安捷伦 B1530A 波形发生器/快速测量模块（WGFMU）组成测试平台进行电学性能测量，如图 2.10 所示。使用探针台将待测器件接入测试平台后，通过半导体参数分析仪和 WGFMU 进行信号的发生和接收。

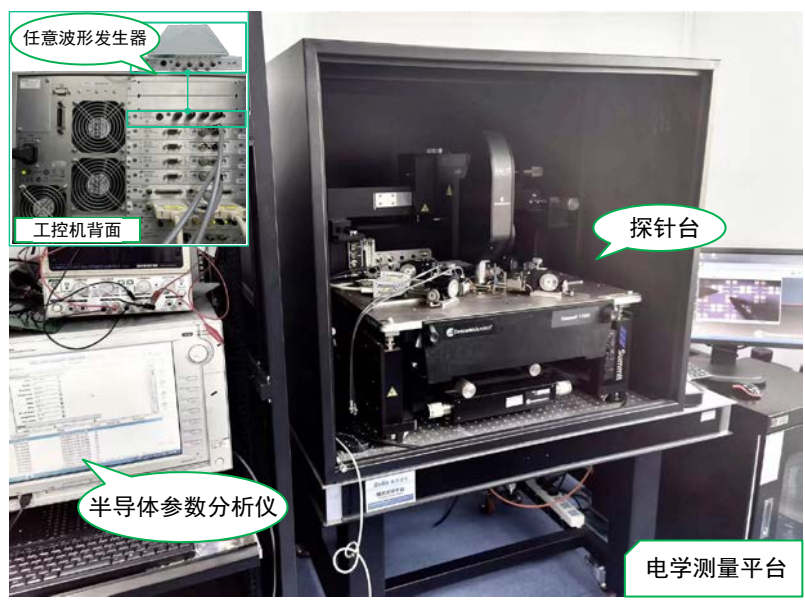

图2.10 电学测量平台

安捷伦 B1530A WGFMU 是一款能够实现任意线性波形生成（ALWG）和同步快速 I/V 测量的仪器，原理图如图 2.11 所示。每个 B1530A WGFMU 通道都可以在快速 I/V 模式和脉冲模式下独立运行。在快速 I/V 模式下，通过 ALWG 函数创建任意波形，并且可以测量电流和电压。在脉冲模式下，可以产生比快速 I/V 模式下更窄的脉冲，并且可以测量电流和电压。在进行电流测量时，B1530A WGFMU 的分辨率为 2nA，采样速度最快可达 5ns。这使得它能够对器件的静态和动态特性进行准确的研究，从而可以更全面地了解器件的性能。

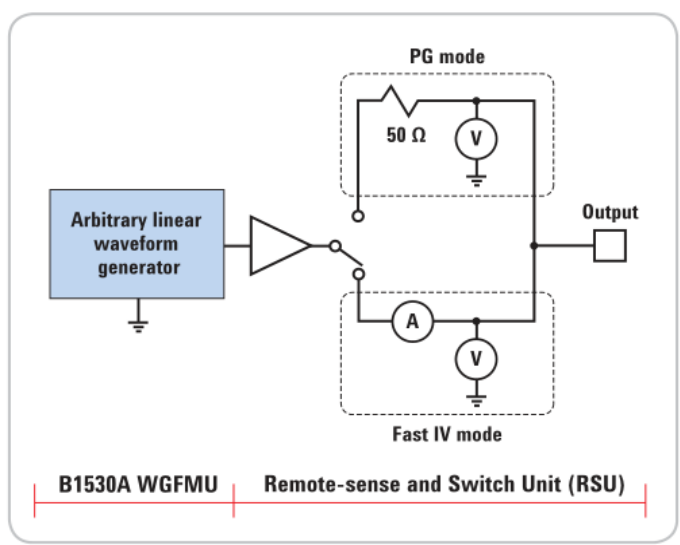

图2.11 B1530A WGFMU 工作原理

Keysight B1500A 半导体器件参数分析仪是一款功能全面的器件表征解决方案。该仪器支持 I/V、C/V 以及脉冲/动态 I/V 测量，能够提供高度可靠的测量结果。除此之外，该仪器还能够捕获其他常规测试仪器无法实现的超快速瞬态现象，并能检测 1kHz 至 5MHz 范围内的多频 AC 电容测量。通过 PC 上的 EasyEXPERT group 中的应用程序控制，可以方便地完成电学测试。在进行测试之前，需要将器件与探针台连接，并对 cascade 探针进行极性定义，软件控制界面如图 2.12（a）所示。然后，可以进行脉冲测量以及伏安特性测量，软件控制界面如图 2.12（b）、（c）所示。以上测量有助于全面地研究器件的静态特性和动态特性。

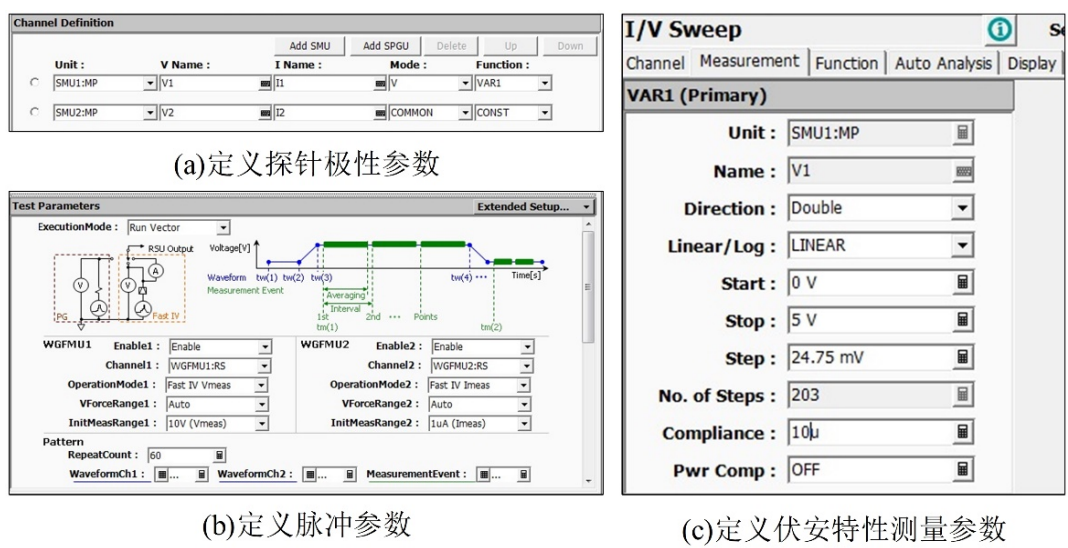

图2.12 电学性能测试软件控制界面（a）定义探针极性参数 （b）定义脉冲参数 （c）定义伏安特性测量参数

## 2.3.2 器件接入测试平台方法

将 Cascade 探针台连接器件到测试平台进行测量。Cascade 探针台能够满足飞安级漏电流测试，实现 10000V 电压的测量，并能完成 500G 器件测量。探针台采用微孔吸附模式固定器件，器件内部发生阻变效应的区域仅限于中间十字交叉的工作区域。由于该区域面积非常微小，因此与探针台的探针直接相连较为困难。为此，探针台的探针都会压在引出的大电极区域上。图 2.13 展示了器件的宏观和微观实物图。

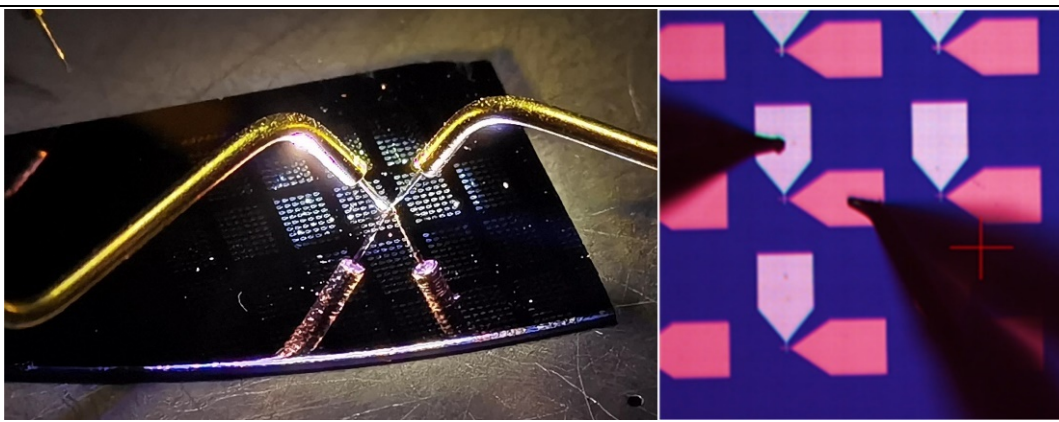

图2.13 器件宏观和微观实物图

在连接器件时，Au 电极的探针连接电源正极，而 Pt 电极的探针则接地。示意图如图 2.14 所示。

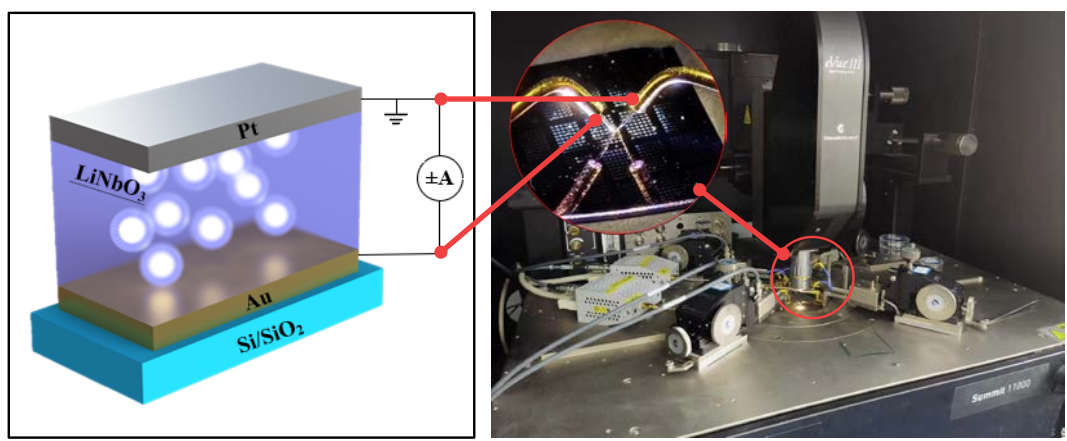

图2.14 器件接入测试平台方式

## 2.4 本章小结

本章首先简要介绍器件的总体结构设计，随后通过对比三种曝光技术的优缺点，选择光学曝光将器件的结构设计和图形信息转移到基片上。然后结合薄膜材料的特性和工艺过程难易，采用 PVD 物理气相沉积的方法来沉积薄膜材料。最后，成功制备出 5 个不同尺寸的 LNO 忆阻器，并详细描述了测试平台对器件的电学测量方法和仪器。接下来，第 3 章将用以上测试仪器和测试方案对器件进行的电学测试。

# 第 3 章 LNO 忆阻器的特性分析

根据忆阻器的伏安特性曲线来研究器件的电学性能和导电机理，并构建适当的导电模型是非常重要的。本章节利用上述制备的 LNO 忆阻器进行电学特性分析，实现以下步骤：首先，测量伏安特性曲线以观察器件的阈值电压和电阻等参数，从而了解器件的性能；其次，通过研究伏安特性曲线的特征，探索器件的导电机理；随后，根据其导电机理构建一个适当的导电模型来描述器件的电学性能；最后，将模型与实验数据进行比较进而验证模型的准确性，以确保模型可以准确地描述器件的电学性能和导电机理。这有助于更好地理解器件的特性，并为后续对 LNO 忆阻器应用的研究提供支持。

## 3.1 器件的阻变特性

### 3.1.1 器件的易失性判定

为了探究器件是否存在阻变特性，本论文对器件进行电学测量，测量其伏安特性，观察是否存在非线性，根据测量结果判断器件为哪种类别的忆阻器。本小节的被测对象为上述制备的尺寸为 $2\times2\mu m^2$ 的器件。测量 LNO 器件的伏安特性，测量结果如图 3.1 所示。

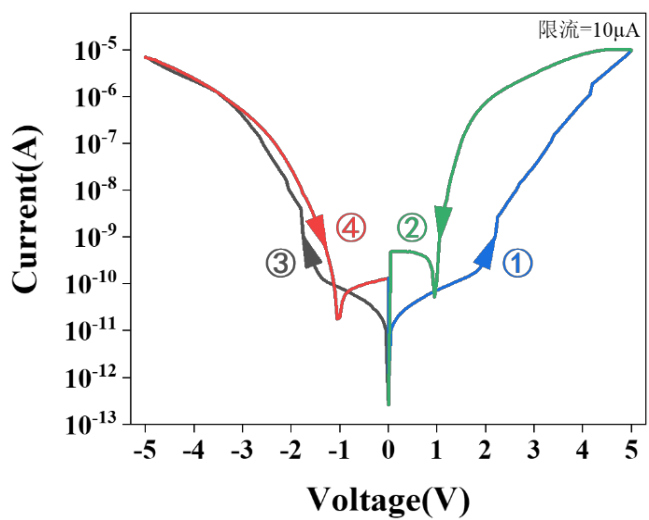

图3.1 器件的伏安特性曲线

首先，如①所示对器件施加正向电压扫描，电压从 0 升压至 5V，再如②所示从 5V 降压至 0；随后，对器件施加负向电压扫描，如③所示电压先从 0 升压至-5V，再如④所示从-5V 降压至 0。本文设定，测量器件的伏安特性需要在 10μA 的限流下进行，以防止器件击穿。器件 HRS 和 LRS 的电阻值一般均选取扫描电压的中值对应的电压电流计算。本次实验扫描电压为 5V，因此根据 2.5V 和-2.5V 处电压对应的电流值计算器件 HRS 和 LRS 的电阻值。图 3.1 中，器件的电流与电压呈非线性关系，初始状态为 HRS。0 至 2V 区间，正向电压扫描时电流上升梯度较小，直至电压超过阈值电压 2V 时，电流梯度陡然升高，器件由 HRS 转变为 LRS，电流在 5V 时达到最大。负向电压扫描时，器件状态为 HRS，电流在-5V 时达到最大。随着电压逐渐减小，器件从 LRS 自发地衰减到 HRS。器件在正、负向上 HRS 和 LRS 的电阻值比分别为在 100 和 2。由于负向电阻比正方向小，后续测量采用正方向的电学测量结果。

器件在正向电压扫描时，从初始的 HRS 转变为 LRS，而在去除电压的短时间内没有保持在 LRS，而是自发衰减恢复到了 HRS。据第 1 章所述，忆阻器的判别准则是根据其去除电压后能否保留 LRS，分为易失性忆阻器和非易失性忆阻器。据此判别出 LNO 忆阻器为易失性忆阻器。

### 3.1.2 器件的稳定性判定

接下来对器件进行多次伏安特性曲线测量，观察其阻变现象，分析器件在 HRS 和 LRS 之间循环转换的稳定性。实验中，对器件施加 5V 电压，充分测量 50 次，期间的伏安特性曲线如图 3.2 所示。当超过 2V 的阈值电压时，器件均从 HRS 转变为 LRS。50 次循环测量得到的伏安特性曲线基本重合，这进一步证明了器件具有良好的稳定性。

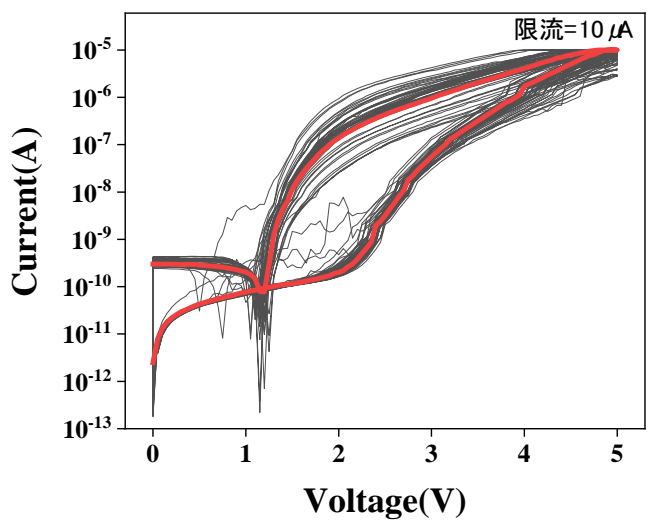

图3.2 器件 50 次循环测量的伏安特性曲线

## 3.2 器件的阻变机理

### 3.2.1 器件的阻变特性分析

根据第一章的论述可知器件的阻变机理分为离子效应、热效应和电子效应。通过分析器件伏安特性曲线的变化，总结器件电阻值的变化趋势能够推导出阻变机理。本小节的被测对象为上述制备的 5 个不同尺寸的器件，其尺寸对比如下图 3.3 所示。

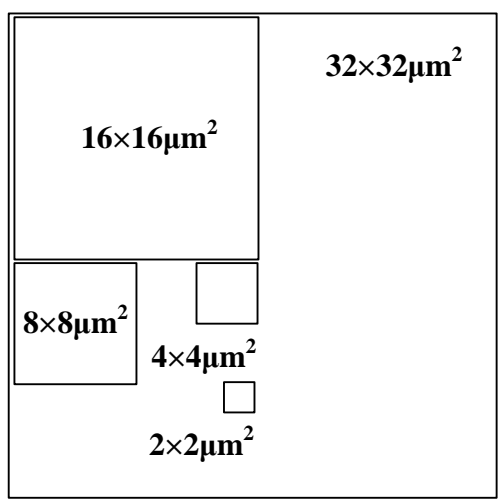

图3.3 5 个尺寸器件的对比

对每个器件都进行伏安特性曲线测量，结果表明五个器件都具有相似的阻变特性，如图 3.4 所示。虽然其阈值电压略有不同，但这些器件不仅具有易失性，而且均存在电阻变化的现象。

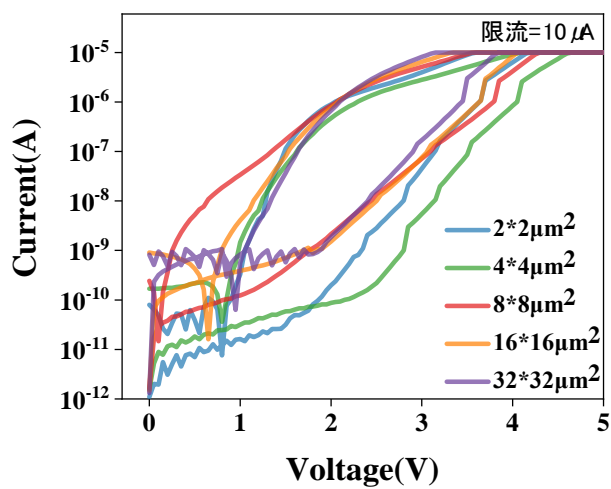

图3.4 5 个器件的单向伏安特性曲线

将 5 个器件的伏安特性曲线在 2.5V 电压处 HRS 和 LRS 的电流值进行对比，结果如图 3.5 所示。五个器件在 HRS 时的电流值会随着器件有效面积的增大而升高，而其在 LRS 时的电流值基本不变，这表明在大电压控制下的器件电流变化主要是由器件内部导电细丝的形成过程决定的。

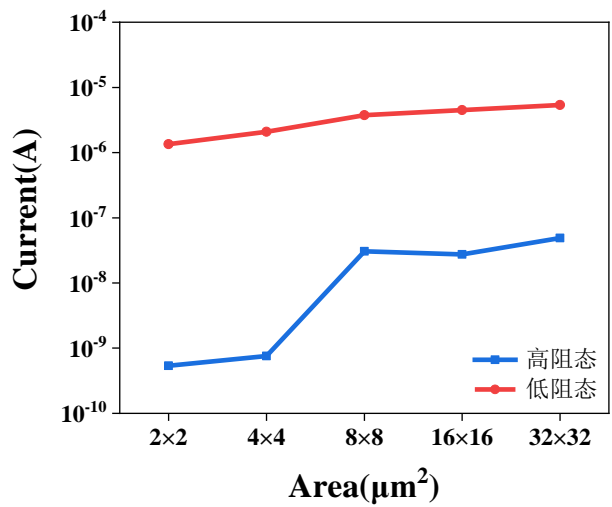

图3.5 5 个器件 HRS 和 LRS 的电流对比

接下来，进一步解释器件内部导电细丝的形成和断裂导致器件电阻变化的原理。器件的初始状态为绝缘态，在未施加电压时，离子运动处于动态平衡的状态，如图 3.6 所示。当给器件施加电压时，在电场的作用下离子做定向移动，阴离子向 Au 电极移动，阳离子向 Pt 电极移动，并在其内部形成导电细丝，此时器件由 HRS 变为 LRS，过程如图 3.6 所示。当撤去器件的外加电压时，离子做扩散运动，导电细丝会自发地断裂，此时器件由 LRS 变为 HRS，过程如图 3.6 所示。综合以上分析本论文可以利用导电细丝机制来解释 LNO 忆阻器的阻变机理。

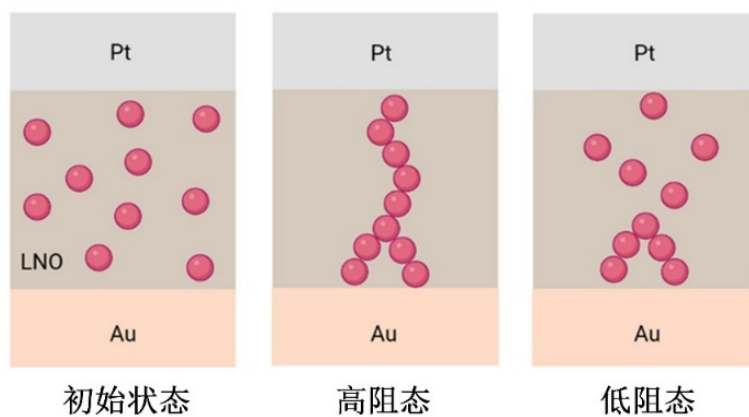

图3.6 器件内部导电细丝的形成和断裂过程

## 3.2.2 器件的阻变机理解释

根据对器件的阻变特性分析可以得知器件的阻变机理是由离子效应中导电细丝的形成和断裂决定的，下面通过分析伏安特性特性曲线进一步对器件的阻变机理进行阐述。本小节的被测对象为上述制备尺寸为 $2×2\mu m^2$ 的器件。其正向的伏安特性特性曲线分为四个过程，对应于图中的①、②、③和④，如图 3.7 所示。

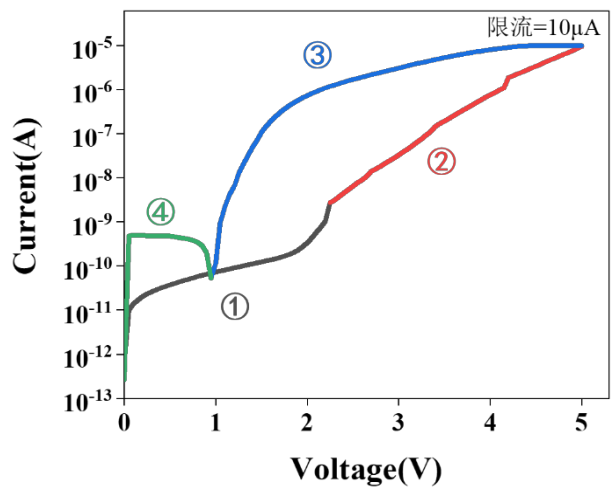

图3.7 器件单次测量的伏安特性曲线

在电场作用下器件内部能做定向移动的带电粒子叫载流子。当给器件施加的电压较小时，电场提供的能量不足以使载流子越过肖特基势垒，器件处于高电阻低电流的状态，对应①过程。随着电压的逐渐增大，载流子越过势垒，离子开始在内部进行定向移动。带正电荷的阳离子向 Pt 电极移动，并在 LNO/Pt 界面附近漂移和积累；带负电荷的阴离子向 Au 电极移动，在 LNO/Au 界面附近漂移和积累，器件内部离子移动方向如图 3.8 所示。载流子的定向移动促成器件内部形成导电细丝，使得器件转换为 LRS，电流逐渐增大，对应②过程。在电压回扫的过程中，虽然电压在逐渐减小，但③过程的电流仍然比②过程的高，这是因为在②过程中形成的导电细丝仍然存在，在此基础上继续施加电压会使得器件内部不断形成更多的导电细丝，所以③过程器件的电流仍然比②过程器件的电流大。

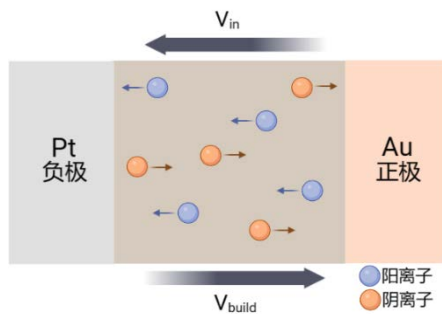

图3.8 器件内部离子迁移过程

器件的外加电场方向是从 Au 电极向 Pt 电极，与此同时器件内部的阴离子和阳离子会发生重新分配，形成内建电场 $V_{build}$，其方向与外加电场 $V_{in}$ 方向相反，如图 3.8 所示。当电压降低到 1V 时，器件的电流最小，对应于图 3.8 中③过程和④过程的交叉点，这是因为此时外加电场 $V_{in}$ 和内建电场 $V_{build}$ 的大小均为 1V，二者处于相对平衡的状态。随着外加电压减小到 1V 以内，器件无法导通。但是由于此时内建电场和离子浓度梯度存在，导致 LNO/Au 和 LNO/Pt 界面附近的离子进行扩散运动，④过程的电流有小幅度的增加。因此，LNO 忆阻器的阻变现象主要是受导电细丝和内建电场的影响。

### 3.2.3 调控器件的阻变现象

为了进一步了解 LNO 忆阻器的阻变原理，本论文进行了更详细的电学测量以展示如何调控导电细丝在器件内部的形成与断裂。本小节实验均采用尺寸为 $2×2\mu m^2$ 的器件进行电学测量。对器件施加不同的正向截止电压，探究其调控器件电阻变化的能力，测量其伏安特性进测量结果如图 3.9 所示，其截止电压分别为 3V、3.3V、3.6V、3.9V、4.2V、4.5V。

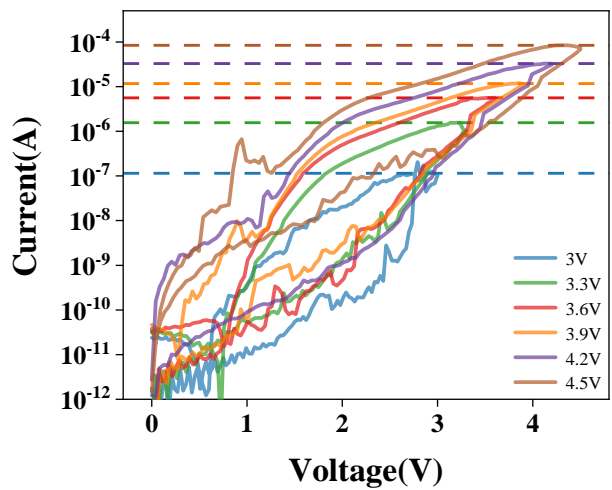

图3.9 不同截止电压下的伏安特性曲线

在 2.5V 电压时对比器件的 HRS 和 LRS 的变化。根据实验结果可以观察到，器件在 LRS 的电流响应随着截止电压的升高而逐渐增大。这种现象可以说明，在外加电场的作用下，随着电场强度逐渐增加，器件内部的离子进行迁移运动的速度和数量增多，使得在两个电极之间形成更多的导电细丝通道，导电细丝的连接强度也随之增加，所以可以观察到器件 LRS 的电流响应逐渐增加。

接下来探究不同脉冲幅值对器件阻变特性的调控。对器件施加频率相同、幅值不同的电压脉冲序列，脉冲频率为 333Hz，其中脉冲持续时间为 1ms，脉冲间隔为 2ms，如图 3.10（a）所示，该过程需要对器件施加连续 10 个脉冲序列进行观测。

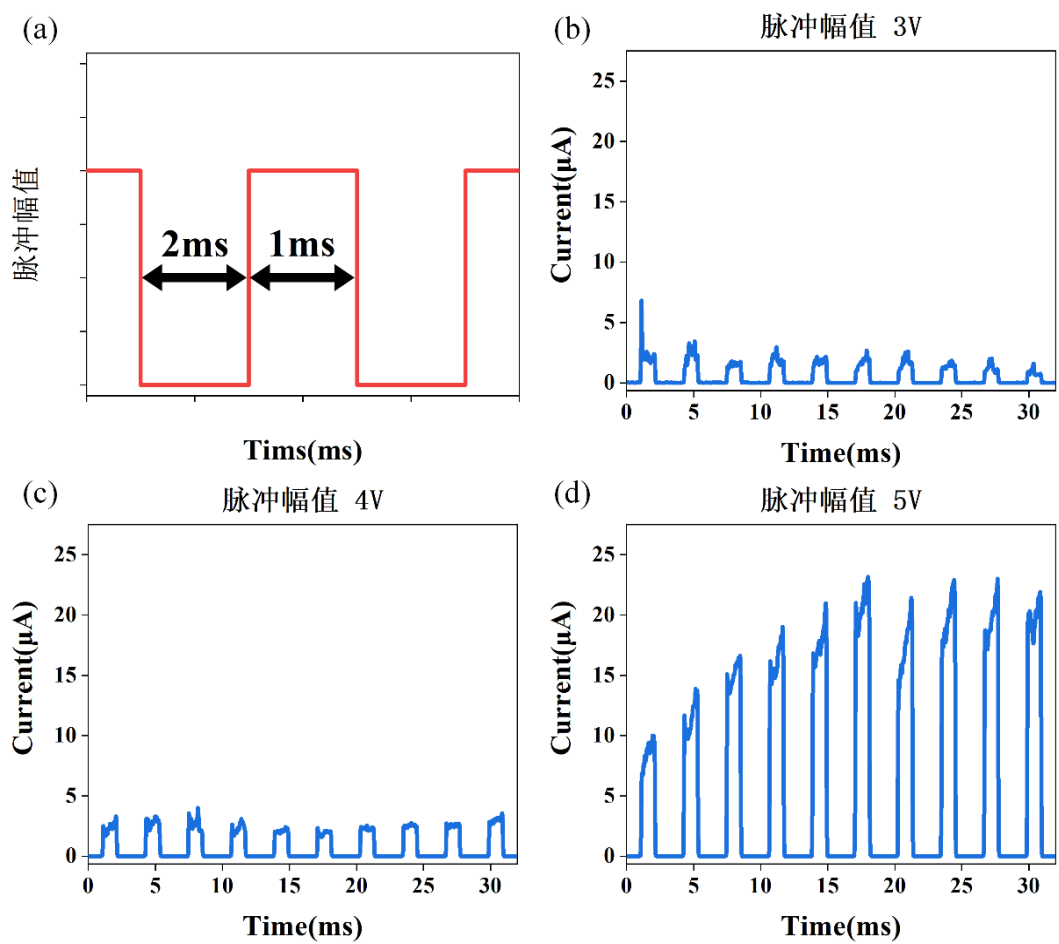

图3.10 不同脉冲幅值的器件响应（a）脉冲时间示意 （b）脉冲幅值为 3V （c）脉冲幅值为 4V （d）脉冲幅值为 5V

对器件施加幅值为 3V 的脉冲序列，其测量结果如图 3.10（b）所示。此时电流基本保持不变，这是因为在小电压下器件内部不足以形成稳定的导电细丝，器件仍处于 HRS，无法观察到电流增大的现象。对器件施加幅值为 4V 的脉冲序列，其测量结果如图 3.10（c）所示，该电流响应有小幅度增强，这是因为在外加电压的控制下已经有导电细丝形成，但它们在脉冲间隔内会自发地断裂。对器件施加幅值为 5V 的脉冲序列，其测量结果如图 3.10（d）所示，此时电流响应明显增强，在几个脉冲后电流达到饱和状态，这是由于在强电压的作用下器件内部形成稳定不易断裂的导电细丝。因此，脉冲幅值和器件电流响应呈线性关系，较高的脉冲电压能够使器件的电流响应发生明显变化。

探究不同脉冲频率对器件阻变特性的调控。对器件施加不同频率但幅值相同的电压脉冲序列，脉冲电压为4V，脉冲时间统一为1ms，而脉冲间隔τ不同，如图3.11（a）所示，在 300ms 内观测器件的伏安特性曲线。

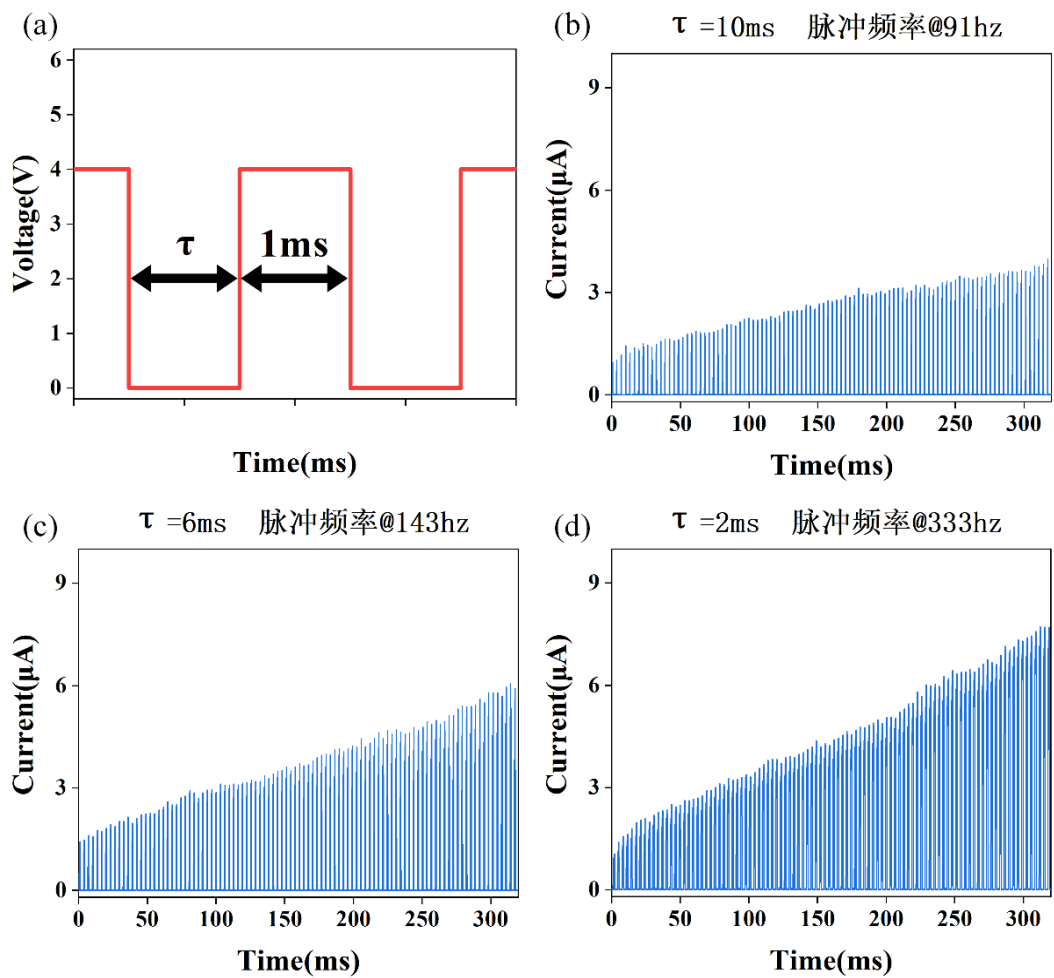

图3.11 不同脉冲频率的器件响应（a）脉冲时间示意 （b）脉冲频率为 91hz （c）脉冲频率为 143hz （d）脉冲频率为 333hz

对器件施加频率 91Hz（τ 为 10ms）的脉冲序列，其测量结果如图 3.11（b）所示。此时电流的增强幅度较小，这是因为较低的脉冲频率会限制器件内部的导电细丝的生长，在下一个脉冲到达之前，刚形成的导电细丝会自发断裂，所以电流变化不明显。保持脉冲幅值不变，对器件分别施加脉冲频率为 143Hz（τ 为 6ms）和 333Hz（τ 为 2ms）的脉冲序列，其测量结果如图 3.11（c）、（d）所示。此时电流随着每一个脉冲的累加而线性增强，并且在幅值为 333Hz 的脉冲测量结果中这种线性关系更加明显。在高脉冲频率的调控下，电流响应随着每一个脉冲的累加而增强，这表明器件从 HRS 转换为 LRS 是动态耦合的。

上述测量结果表明较高的脉冲幅值和脉冲频率可以使得器件的电流响应变化明显增强。这是因为在有脉冲电压的情况下，器件内部的导电细丝逐渐生长，器件的

电阻逐渐降低，器件由 HRS 变为 LRS；在有脉冲间隔到来的情况下，虽然导电细丝的生长停止逐渐衰减，但高脉冲电压和高脉冲频率使得在下一个脉冲达到之前导电细丝仍然存在并没有完全断裂，即导电细丝的生长并不是从零开始而是从当前器件内残余的导电细丝的基础上开始的，使得器件的电流是逐渐增加的。这表明器件的每一时刻电流响应不仅与当前时刻的电压输入有关也与前一刻器件的电流状态有关，这种电流响应的变化是动态耦合的。

## 3.3 本章小结

综上所述，通过对器件进行伏安特性测量并分析测量结果，表明 LNO 忆阻器为易失性忆阻器，并且兼备稳定性。为了进一步探究器件的阻变机理，对 5 个不同尺寸的器件进行伏安特性测量。结果表明，器件的阻变特性主要是导电细丝和内建电场的影响。最后建立了一种导电细丝机理的导电模型，并发现器件内部导电细丝的形成和断裂是可调控的，较高的电压强度和脉冲频率都可以使器件内部形成稳定的导电细丝，电阻变化的也比较明显。

# 第 4 章 LNO 忆阻器的应用

经过上述 LNO 忆阻器的电学测量结果分析，表明该器件具有优异的非线性和短期记忆特性，因此非常适合作为储备池系统的非线性节点。本章提出了一种基于 LNO 忆阻器的 RC 系统，利用器件内部动态的物理过程可以获得丰富的储层状态，并将其表示为忆阻器的输出电流组合，然后通过简单的读取功能进行后续处理。本论文使用该系统进行实验，并成功地完成了对标准阿拉伯数字 0-9 的识别任务。

## 4.1 基于 LNO 忆阻器的 RC 系统结构设计

### 4.1.1 系统结构

利用 RC 处理时间序列信号非常有效，由于储层的状态不仅受当前输入的影响，还受历史输入的影响，因此非线性节点必须具备短期记忆的能力。上一章测试结果表明，LNO 忆阻器具有短期记忆的特性，该器件的输出电流值不仅由当前输入电压决定，还与前一刻的状态有关。此外，忆阻器具有的非线性特性可以将输入空间的数据点映射到高维特征空间，在该空间中的输出能够线性可分，这更有利于后续的信息分类处理。所以将 LNO 忆阻器作为 RC 储层单元是值得研究的。

本论文构建 LNO 忆阻器的储备池计算系统，进行标准阿拉伯数字 0-9 的识别任务，系统结构包括输入层、储层、输出层三部分，如图 4.1 所示。

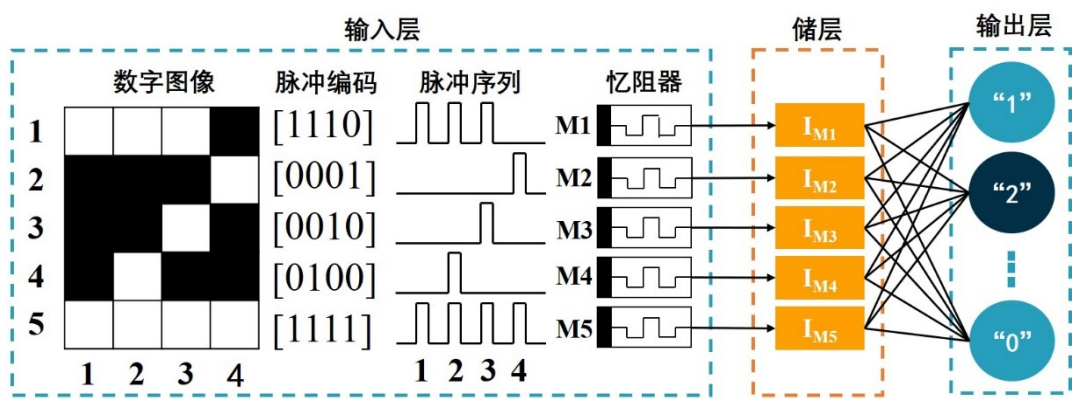

图4.1 储备池计算系统的三层结构

输入层由半导体分析仪作为信号发生和接收装置，储层由五个 LNO 忆阻器并联组成，输出层是通过 Python 进行训练和输出结果。在输入层，首先将 5×4 数字图像

转换为二进制脉冲编码，通过自定义将脉冲编码转换为电压脉冲序列，由半导体分析仪输入到 LNO 忆阻器。储层中的器件对输入的电压脉冲序列进行非线性变换，储层的状态由五个器件最后一个脉冲的电流值组合表示，并通过半导体分析仪记录。储层将预处理后的输入信息非线性地映射到高维特征空间中，使得输出信号变得线性可分，随后对其增加一个线性网络，可以对输出进行分类和处理，即可完成任务。输出层包含两个过程，训练和识别过程。训练过程是训练输出层的 5×10 的权重矩阵，这个矩阵是通过 Moore-Penrose 伪逆求出来的，利用储层的储备池输出向量和训练样本标签的向量求伪逆矩阵，训练矩阵的目的就是把储层的输出拟合成相应的输出类别即数字 0-9。识别过程是把测试样本输入到储层中，储层的电流响应作为已经训练好的矩阵的输入，得到的输出就是分类的结果即数字。

## 4.1.2 动态特性

传统 RC 的储层是由许多互相连接的节点组成，量级在 $10^2$~$10^3$ 之间，状态可通过网络读取。而基于 LNO 忆阻器的 RC，凭借其短期记忆特性，可以实现不同时间步信号之间的相互耦合，一个器件可替代多个节点组成的环路。器件对输入信号响应时，内部存在多个虚拟节点，虚拟节点的状态由当前输入和前一刻的状态决定，它们之间是非线性耦合的。忆阻器具有非线性特性和短期记忆，器件受电激励后的内部动力学过程对应着各虚拟节点的状态，器件的输出与当前输入和前一刻的状态有关，因此适合处理时间序列信号。

在本论文的RC系统中，忆阻器接收从输入层传输的脉冲序列，最后一个脉冲的电流响应表示为储层的状态，前三个脉冲对应的电流响应即为虚拟节点的状态（如图 4.2 所示）。虚拟节点不需要其他器件或者其他输出节点，仅通过一个器件就可以包含虚拟节点的全部信息，即其前三次的记忆会保留在器件中，并体现在最后一个输出电流值。而且虚拟节点隐藏在忆阻器的响应中，所以只需要训练输出层，而不需要训练储层。因此，该 RC 系统的工作效率会整体提高。

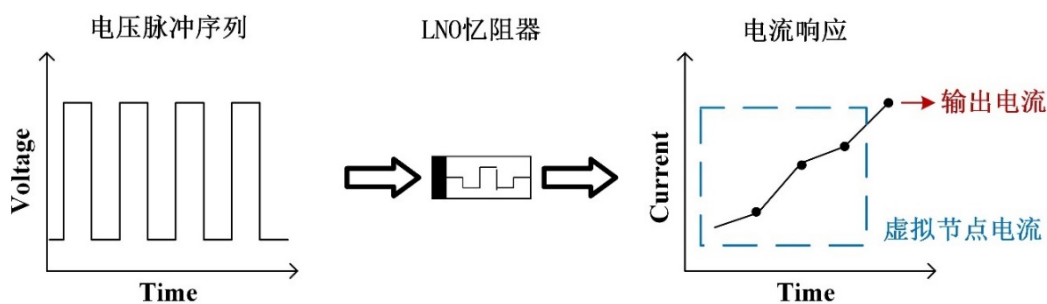

图4.2 器件电流响应示意

当忆阻器用于 RC 系统时，其非线性特性可以通过对系统的动态行为进行调节和控制，从而实现提高系统的性能和功能。相比于线性系统，忆阻器的非线性特性能够为系统提供更多的自适应和记忆能力，使系统能够适应更加复杂的环境和信号输入，从而具有更好的稳定性和鲁棒性。同时非线性特性还可以增加系统的计算能力，提高系统的处理速度和准确性，使其更适合进行神经形态计算和人工智能应用等任务。而另一方面，忆阻器的易失特性使得其状态可以被外界信号或者电压所改变。这种特性可以被用于模拟神经元的兴奋和抑制过程，同时也可以用于实现动态储备池计算。在忆阻器的易失特性的作用下，系统的状态可以根据输入信号的变化而发生快速变化，这种变化可以被用于实现各种计算和信号处理的任务。此外，忆阻器的易失特性还可以在系统出现故障时对系统进行快速重置，从而增加了系统的可靠性和稳定性。

## 4.2 基于 LNO 忆阻器 RC 系统的数字图像识别任务

### 4.2.1 系统的输入层

本章利用 LNO 忆阻器的 RC 系统，演示了数字识别任务的处理过程。该任务的输入是机写 0-9 的数字图像，每个图像由 5×4 的像素格子组成，每个格子有黑白两种状态，如图 4.3 所示。

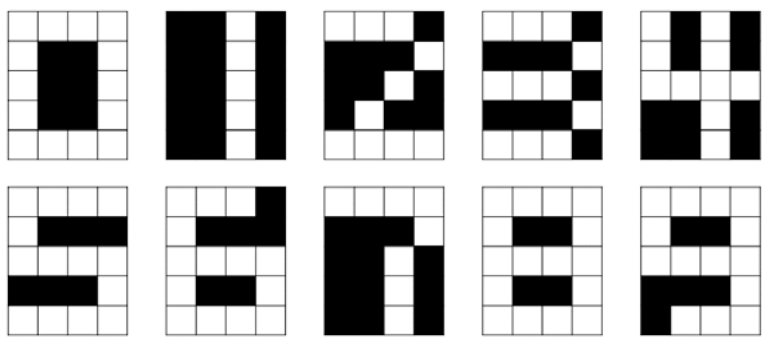

图4.3 机写 0-9 数字图像

将黑白色的图像转换成电压脉冲序列要使用时分复用策略。该方法将图像中的每一个像素看作是一个输入通道，通过对每个输入通道进行时分复用，可以将每个像素转换成一个时间序列上的电压脉冲。具体来说，对于每一个像素都会被分配一个时间窗口，在该时间窗口内，如果像素是白色，则产生一个电压脉冲，如果像素是黑色，则不产生电压脉冲。通过这种方式，可以将整张图像转换成一个时间序列上的电压脉冲序列，该序列可以作为 RC 系统的输入。

首先,将白色像素格定义为"1",黑色像素格定义为"0"。然后将 5×4 的数字图像分成 5 行,每行包含 4 个像素格的信息。这样,数字图像信息可以转化为 5 个二进制脉冲编码,每个脉冲编码包含 4 个二进制数字,这 4 个数字按照从左到右的顺序对应于每行像素格的状态。例如,数字图像 2 的第一行像素格状态为白、白、白、黑,对应的数字脉冲编码为[1110]。共有 10 种数字图像的二进制脉冲编码,分别为[1000]、[0100]、[0010]、[0001]、[1010]、[1001]、[0110]、[1110]、[0111]和[1111]。

然而,二进制脉冲编码无法直接传输到器件中,需要将其转换为电压脉冲序列。每行像素格包含 4 个像素信息,因此每个电压脉冲序列包含 4 个尖峰电压。将二进制编码为 0 的电压脉冲定义为低电平 2.5V,将二进制编码为 1 的电压脉冲定义为高电平 5V。例如,数字图像 2 的第一行脉冲编码序列为[1110],对应的电压脉冲序列为 5V、5V、5V、2.5V。

为了确保各个虚拟节点之间的动态耦合,电压脉冲的时间间隔设为 1ms。因为 LNO 忆阻器具有短期记忆,若没有电压激励,器件将回到没有发生阻变的初始状态即 HRS,因此脉冲间隔需要足够短,以防止器件的记忆完全消失。因此,每个 5×4 的数字图像都可以转换为 5 个电压脉冲序列作为 RC 系统的输入信号,如图 4.4 所示。

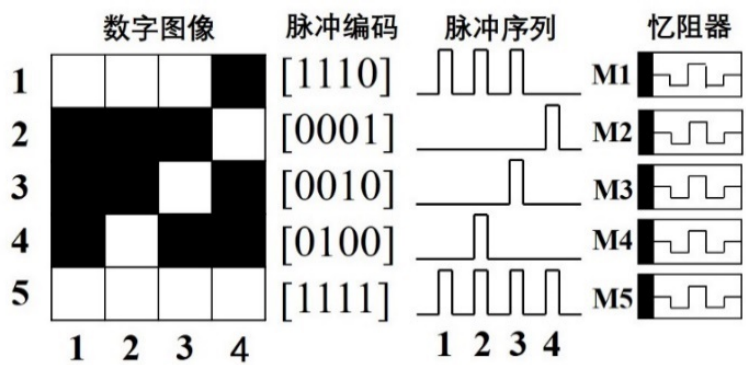

图4.4 RC 系统的输入层

为了有效处理时间序列信号即电压脉冲序列,这要求对输入信号进行响应的器件必须具有短期记忆,因为 RC 不仅需要考虑当前输入的信号还需要考虑历史的输入,LNO 忆阻器可以利用固有的短期记忆和非线性特性来处理时间序列信号。器件的短期记忆意味着,器件当前的输出不仅取决于当前的输入也取决于历史输入的影响,这使得RC系统可以获得丰富的储层状态。器件的非线性特性可以将处理后的电压脉冲信号映射到高维特征空间,对输入的电压脉冲序列进行时空特征提取,用器件的瞬态电流响应代表提取到的特征。这使得忆阻器的输出变得线性可分离,从而可进行分类和回归。

## 4.2.2 系统的储层和输出层

RC 系统中的储层应当有 5 个非线性节点，这是因为输入层利用器件的非线性特性和短期记忆，将数字图像经过时分复用的策略处理成 5 个电压脉冲序列，忆阻器可以将输入的电压脉冲序列映射到高维特征空间中，对输入的电压脉冲进行时空特征提取，用忆阻器的瞬态电流响应代表提取到的特征。储层的状态由 5 个器件提取到的最后一个脉冲的瞬态电流值组合表示。

本论文利用回声状态网络进行输出层的训练和识别过程。由于储层中的忆阻器替代了具有延迟反馈的非线性节点的动力学系统，储备池计算的动态特性由忆阻器本身特有的动力学特征所表示，将输入层的信息映射到高维特征空间中，使得储层的输出变得线性可分，所以只用训练输出层的权值 $W_{out}$ 即可，如图 4.5 所示。

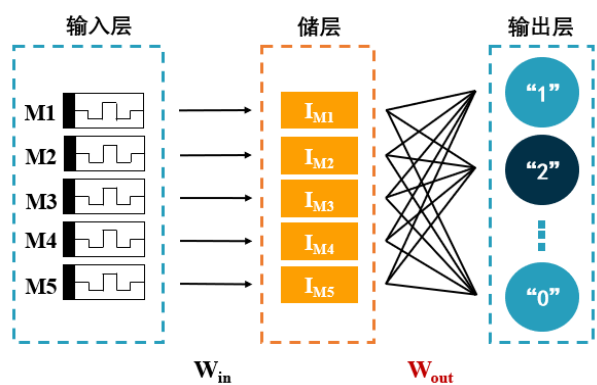

图4.5 RC 系统的输出层

这个网络的输入是 5 个储层的输出向量即 5 个器件最后一个脉冲的电流值，中间层生成一个随输入不断变化的复杂动态空间，利用这些内部状态，线性地组合出所需要的输出即数字 0-9，因此训练算法也就简化为一个线性分类器。

假设定义这个回声状态网络中 $x(k)$ 为储层的状态，$S(k)$ 为第 $k$ 步结束时的输出值。在 $k$ 时间步的输出值依公式（4.1）计算：

$$S(k) = Wx_{(k)} \quad (4.1)$$

利用 0-9 这 10 张数字图像的 20 次测试样本的数据，对回声状态网络进行训练，获得最优权重矩阵。输入是储层 5 个器件的电流值，输出是数字 0-9，这是一个 10 分类的问题，因此这个权重矩阵 $W$ 的规模为 5×10。

为了实现权值的计算，根据训练样本中的储层状态 $x$ 和目标输出 $y$，计算出连接权值 $W$，因为状态变量 $x$ 和系统实际输出 $S$ 是线性关系，需要实现的目标是利用网络实际输出 $S$ 逼近实际输出 $y$。权重矩阵的表达式如（4.2）。

$$W = \left( yS^{\dagger} \right)^{T} \quad (4.2)$$

*S* 为实际输出向量，*y* 为目标输出向量，†表示为 Moore-Penrose 广义逆。

为了避免对训练数据进行过拟合，需要最小化 $\|WS - Y\|^2 + \|\lambda W\|^2$，其中 $\lambda$ 为正则化参数。在训练过程中加入结构误差函数，在正则化参数的同时最小化经验误差函数，最小化经验误差是为了极大程度的拟合训练数据，正则化参数是为了防止过分的拟合训练数据，因此对系数进行一定的惩罚。

将 10 个数字的 20 次测量结果即 200 个数据作为训练集，进行输出层的训练，另取 10 个数字的 10 次测量结果即 100 个数据作为测试集。利用回声状态网络对 RC 系统的输出进行训练和识别，基于 LNO 忆阻器的 RC 系统成功完成了数字识别任务。

## 4.3 数字图像任务测试结果与性能分析

### 4.3.1 单器件的测试结果与性能分析

在 RC 系统中，输入层将图像信息处理成电压脉冲序列，并通过忆阻器得到相应的电流响应。对于数字 0-9 的图像信息，经过编码和分类整理后，共有 10 种不同的电压脉冲序列和相应的电流响应。图 4.6 中所示的是器件对于这 10 种不同的电压脉冲序列的 30 次电流响应的统计结果。

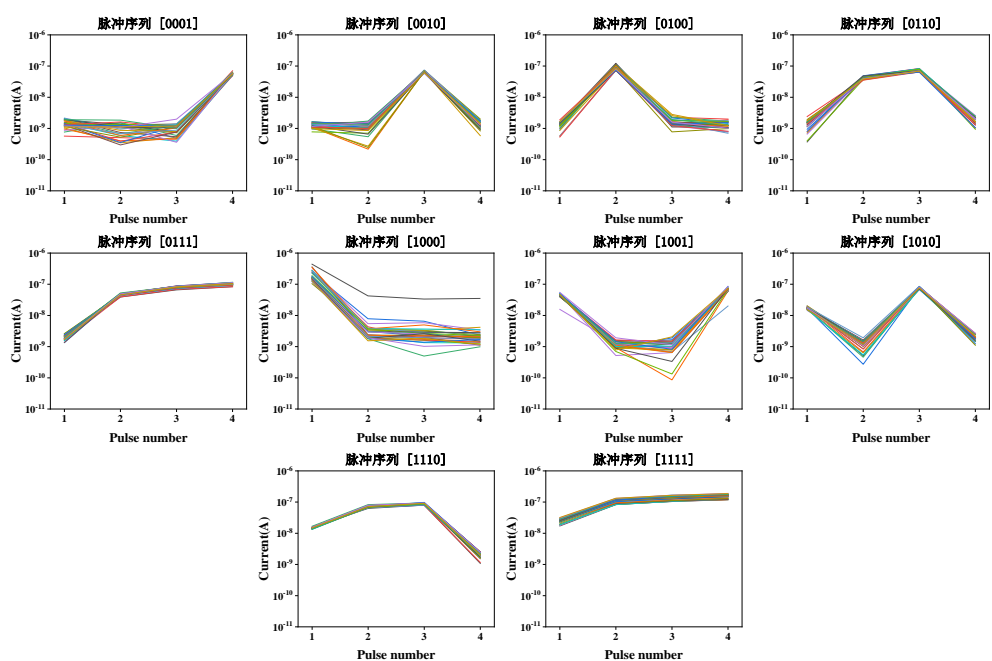

图4.6 10 种不同的电压脉冲序列的 30 次测量结果

第 4 章 LNO 忆阻器的应用

图 4.7 显示了器件对于 10 种不同脉冲序列的 30 次电流响应的均值的标准差的统计结果。标准差整体地反映了数据与平均值的偏离程度，可以根据图中的统计结果看出，器件对不同电压脉冲序列的 30 次电流响应较为平均。该忆阻器的阻变机理是器件在接受电压激励后，内部形成导电细丝，从而使器件从 HRS 变为 LRS，撤去电压激励后，器件会逐渐恢复到 HRS。由于器件的物理特性稳定，因此对于同一种电压脉冲序列的电流响应也相对稳定。

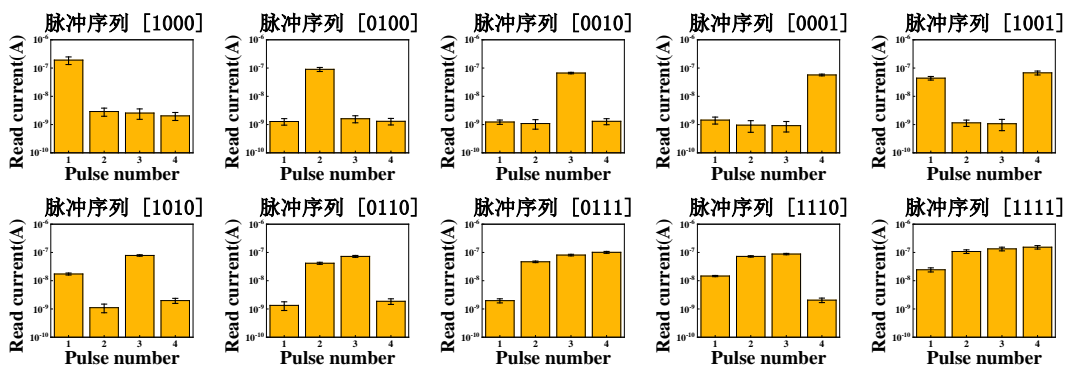

图4.7 10 种不同脉冲序列的 30 次电流响应的均值的标准差

该 RC 系统储层的状态由 5 个器件提取到的最后一个脉冲的瞬态电流值组合表示，但器件在制备的过程中，不同器件的薄膜厚度可能会有细微的差别，也会对器件的电学性能有一定的影响，所以需要对器件的均一性进行研究。本论文选取电压脉冲编码[0100]进行测量研究，比较器件对同一个电压脉冲序列的电流响应变化。由图 4.8 中数据的结果表明器件对于相同电压脉冲序列的电流响应趋势相同。因此，可认为该组器件具有良好的均一性，对同一个脉冲序列，5 个器件的电流响应基本相同，为后续输出层的训练提供稳定的数据支持。

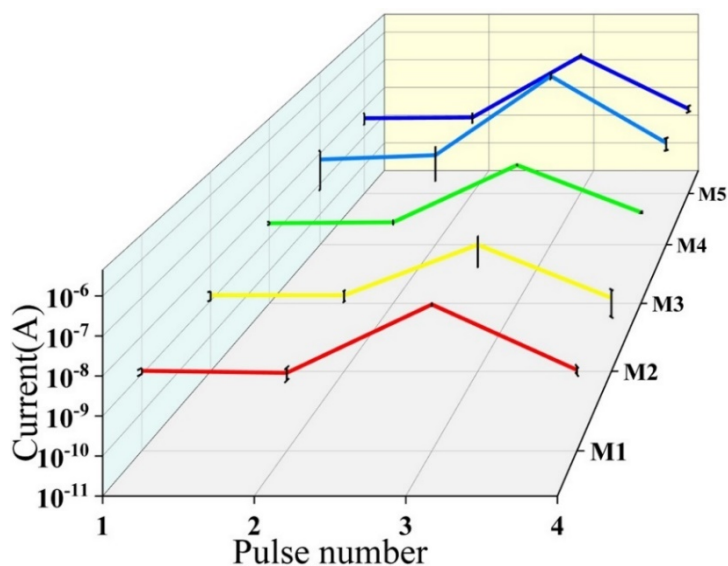

图4.8 经 30 次重复实验，输入脉冲编码[0100]后五个忆阻器的电流均值

## 4.3.2 储层的测试结果与性能分析

RC 系统具有丰富的动态特性，其主要是利用忆阻器的非线性响应和短期记忆。5×4 的数字图像由 20 个黑白像素格子组成，将其转化成 5 个 4 位二进制脉冲编码，再将二进制脉冲编码转换成 5 个电压脉冲序列，每个电压脉冲序列包含 4 个尖峰电压脉冲。忆阻器将输入的电压脉冲信号非线性地映射到高维特征空间，对输入的电压脉冲序列进行时空特征提取，用忆阻器的瞬态电流响应代表提取到的特征。按照"红橙黄绿蓝"颜色排序，依次代表第一个器件到第五个器件的电流响应，如图 4.9 所示。

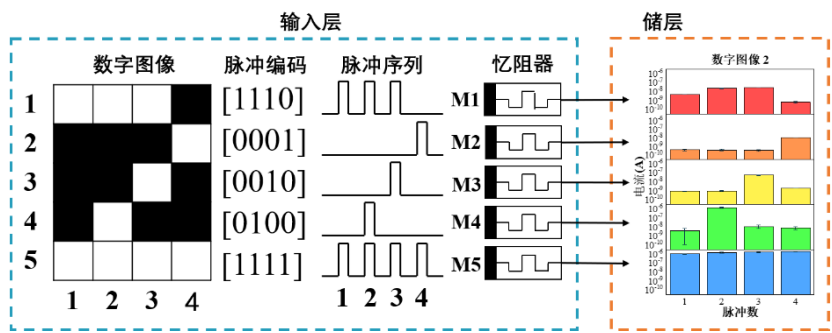

图4.9 储备池计算系统的输入层和储层

图 4.10 显示了 RC 系统在识别 0-9 数字图像时对应的储层状态，图中的误差棒是 30 次测量数据的标准差。

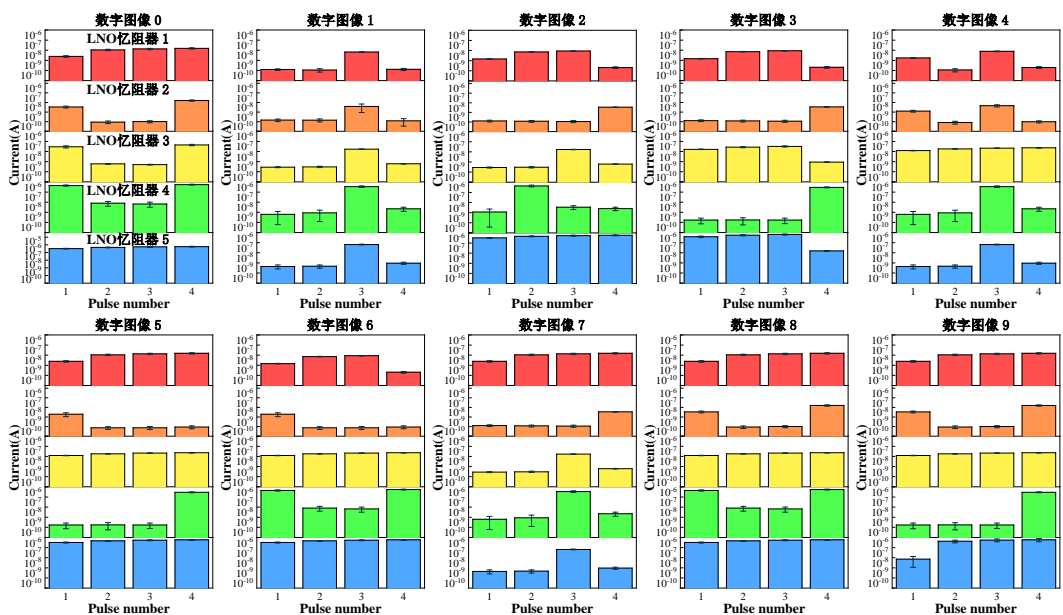

图4.10 0-9 数字图像对应的储层状态

通过直观比较数字"0"、"8"和"9"的图像的储层状态,可以发现它们非常相似。为了证明 RC 系统对于不同数字的图像具有辨别能力,引入统计学中的假设检验方法,通过检验不同储层状态之间是否存在显著性差异来客观、定量地描述系统的模式识别能力。评判该系统对数字图像的储层状态是否有显著性的差异,可以用统计量的 $p$ 值来表示,其值越小则表明模式识别能力越强。求解统计量 $p$ 值需要对 10 个数据集进行多元统计分析,采用 BOX´M 检验和 Hotelling´s T-square 检验的方法。BOX´M 检验是通过检验不同数据集之间的方差是否相等来判断,Hotelling´s T-square 检验是通过比对不同数据集之间的均值差异来判断。这两个检验方法从不同的角度对 10 个数据集进行多元统计分析,可以得出更全面的统计结果,从而验证不同数据集之间是否存在显著性的差异。接下来将每个数字图像的 30 次测量结果表示为数据集 $A_0$、$A_1$...$A_9$,每个数据集为一个 30 行 5 列矩阵,如图 4.11 所示。

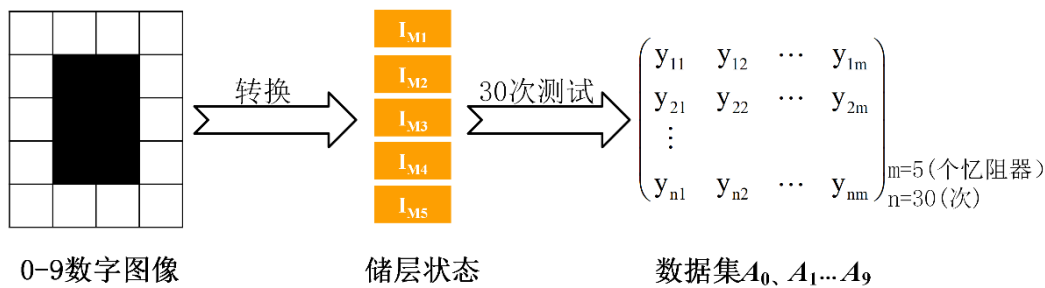

图4.11 数据集组成

BOX´M 检验的步骤为:

(1) 进行假设检验,零假设为数据集 $A_0$、$A_1$...$A_9$ 的方差相等。

(2) 计算每个数据集的样本均值和样本协方差矩阵。

(3) 计算 BOX´M 的统计量 $M$,依公式(4.3)代入计算:

$$M = -2\ln\lambda \quad (4.3)$$

其中,$\lambda$ 为似然比统计量。

$$\lambda = \frac{\prod_{i=1}^{k}|S_i|^{(n_i-1)}}{|S_p|^{(n-k)/2}}, \ i=1,2,\cdots,10$$

$$S_i = \frac{1}{n_i-1}\sum_{j=1}^{n_i}(x_{ij}-\overline{x}_i)(x_{ij}-\overline{x}_i)'$$

$$\overline{x}_i = \frac{1}{n_i}\sum_{j=1}^{n_i}x_{ij},$$

$$S_p = \frac{1}{n-k}\sum_{i=1}^{n_i}(n_i-1)S_i = \frac{1}{n-k}E$$

$$n = \sum_{i=1}^{k} n_i$$

（4）计算 BOX´M 统计量 $M$ 的 $p$ 值，需要将 $M$ 与 $F$ 分布进行比较得出结果。

（5）显著性水平 $\alpha$ 设为 0.05。当 $p$ 值大于或等于 $\alpha$ 时，零假设成立。当 $p$ 值小于 $\alpha$ 时，拒绝零假设，即数据集 $A_0$、$A_1$...$A_9$ 的方差不相等具有显著性的差异。

计算结果如下图 4.12 所示，图中的 $p$ 值均远远小于 $\alpha$，表明不同数据集有显著性的差异，该系统可以很好地区分 10 个数字图像的储层状态。

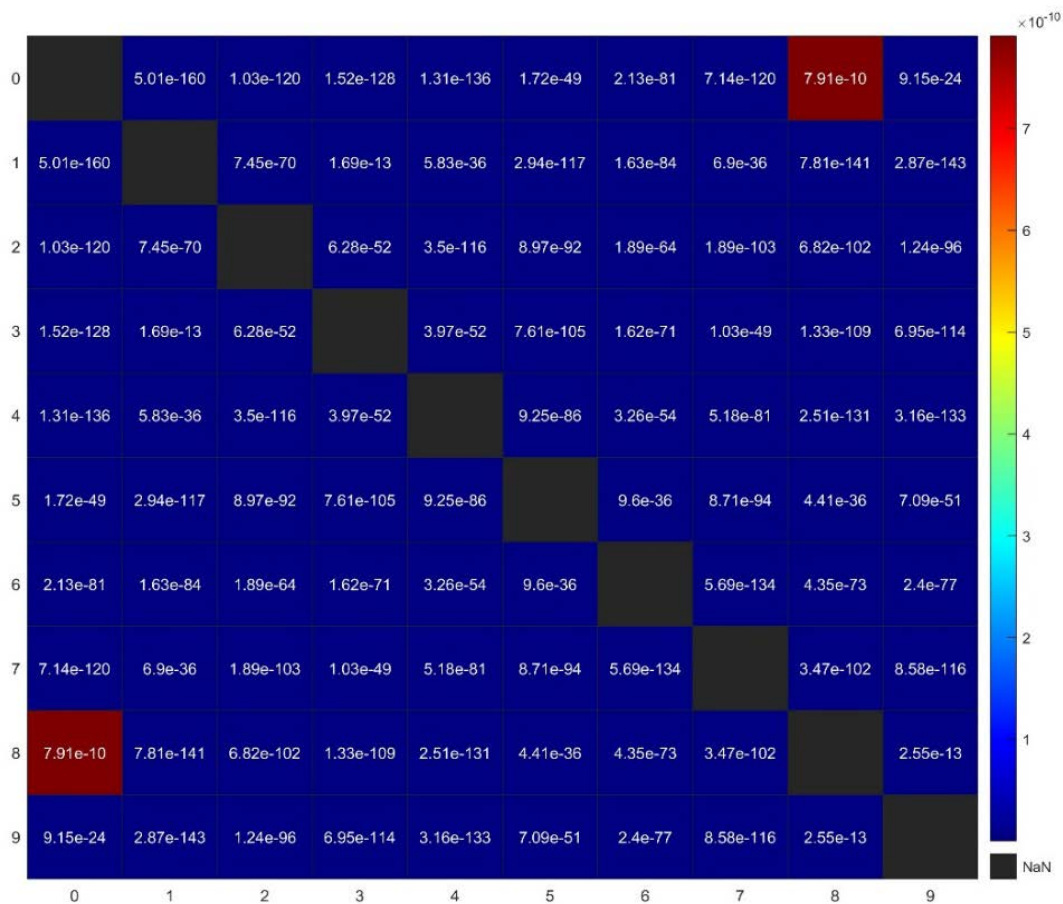

图4.12 0-9 数字图像对应的储层状态的 BOX´M 检验 $p$ 值统计分析结果

注意，图 4.12 中 $p$ 值数据沿正对角线镜像相等，这是一个对称矩阵。每个 $p$ 值代表了对应行和列，两组数字图像的储层状态组成数据集之间的对比结果。例如该矩阵中的 $p(2,5) = p(5,2)$，指图像 "2" 和 "5" 经忆阻器处理后的储层状态的数据集 $A_2$ 和 $A_5$ 的对比结果，示意图如图 4.13 所示。

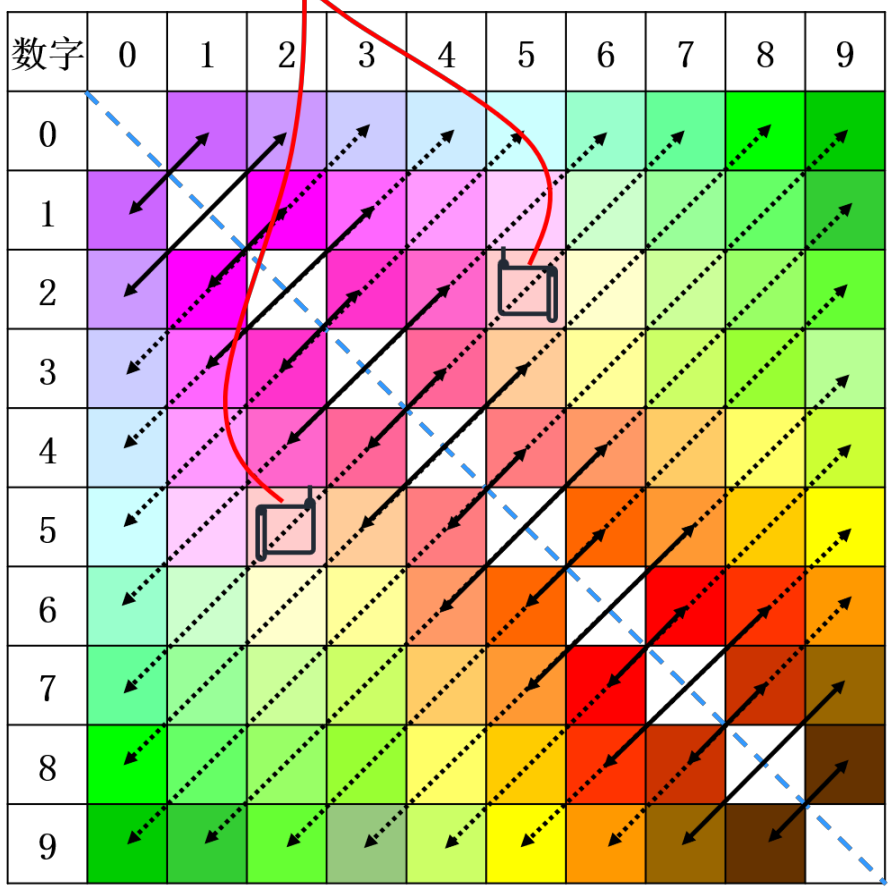

图4.13 $p$ 值数据分布示意图

Hotelling´s T-square 检验的步骤为：

（1）进行假设检验，零假设为数据集总体均值之间没有明显差异。

（2）计算每个数据集的样本均值和协方差矩阵。

（3）计算 Hotelling´s T-square 检验的统计量 $T^2$，依公式（4.4）代入计算:

$$T^2 = n\overline{X}'S^{-1}\overline{X} \tag{4.4}$$

其中，总体均值之差的矩阵转置，S 样本协方差矩阵的逆。

$$\overline{X} = \frac{1}{n}\sum_{i=1}^{n} x_i,\ i=1,2,\cdots,10$$

$$S = \frac{1}{n-1}\sum_{i=1}^{n_i}(x_i - \overline{x})(x_i - \overline{x})'$$

（4）计算 $p$ 值，通过对 $M$ 与 F 分布进行比较得出结果。

（5）显著性水平 $\alpha$ 设为 0.05。当 $p$ 值大于或等于 $\alpha$ 时，零假设成立。当 $p$ 值小于 $\alpha$ 时，拒绝零假设，即数据集 $A_0$、$A_1$...$A_9$ 总体均值之间存在显著差异。

计算结果如下图 4.14 所示,图中的 $p$ 值均远远小于 $α$,表明不同数据集有显著性的差异,该系统可以很好地区分 10 个数字图像的储层状态。

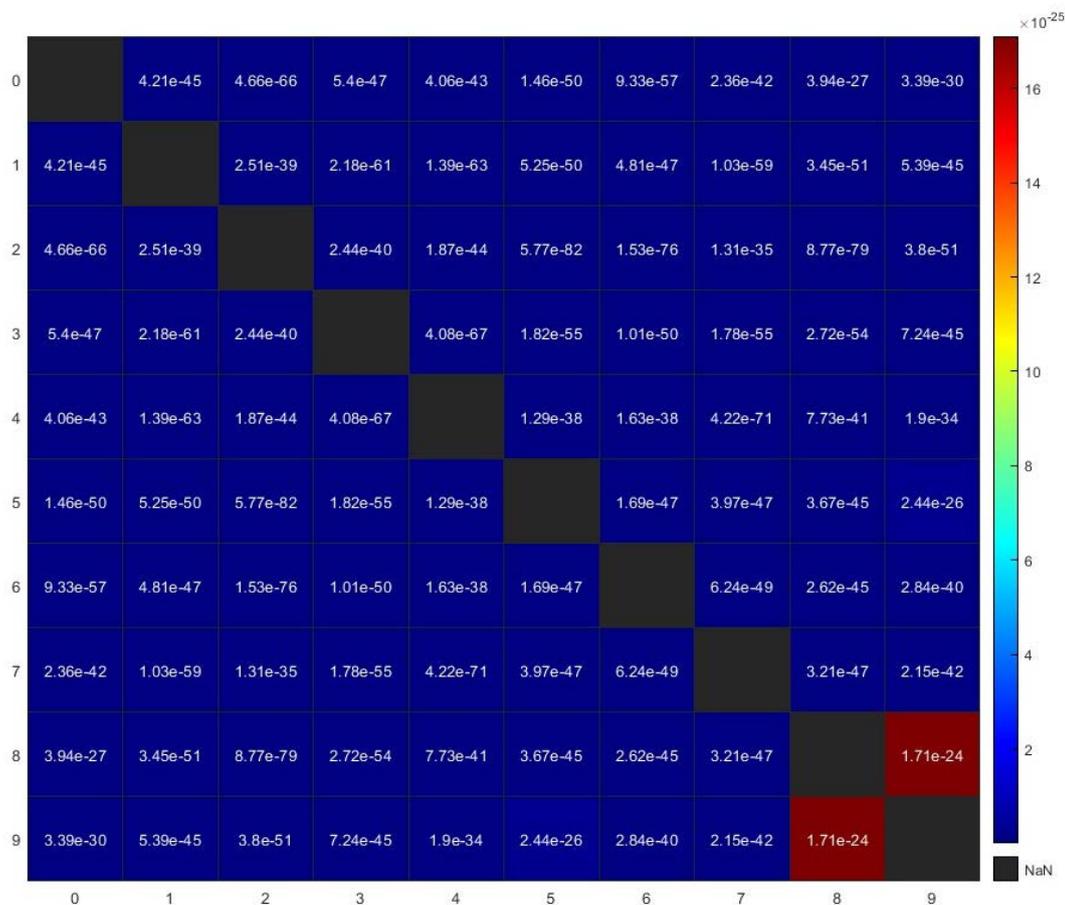

图4.14 具有不等协方差矩阵的 Hotelling´s T-square $p$ 值统计分析结果

相对而言,图 4.12 中的第 8 行第 9 列及其对角线镜像位置的 $p$=7.91e-10;图 4.14 中的第 8 行第 9 列及其对角线镜像位置的 $p$=1.71e-24。这两个值虽然都远小于 0.05,但相对其他值要大一些。$p$ 值远小于 0.05,这就表明数字"8"和"9"通过忆阻器能够很好地识别。但相比于其他配对的数字,区分这两个数字较为困难。用肉眼观察图 4.15 进行模式识别,对比"0"和"8"、"8"和"9"不难看出,每组配对中仅有 2 个像素的差异。统计结果和直观的感受吻合。

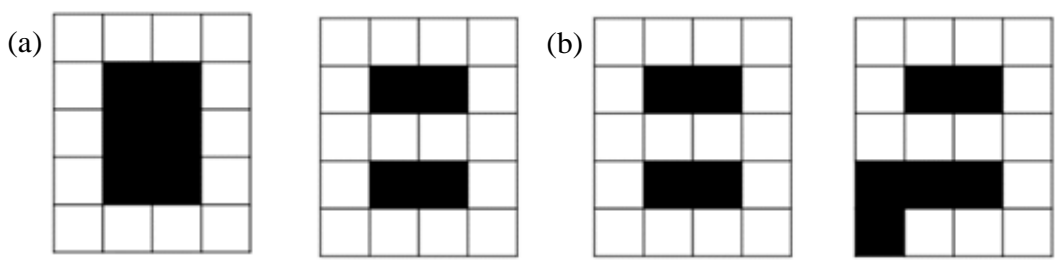

图4.15 两种统计结果指出容易混淆的数字(a)BOX´M 检验表明 0 和 8 容易混淆(b)Hotelling´s T-square 检验表明 8 和 9 容易混淆

## 4.4 本章小结

本章详细阐述了基于 LNO 忆阻器的 RC 系统的系统结构和动态特性，其关键原理是通过单个忆阻器在时间域上形成许多动态耦合点来替代传统 RC 系统中在空间域上形成的延迟反馈环互连节点，器件的非线性响应和短期记忆使得 RC 系统也获得丰富的储层状态。本研究利用 5 个 LNO 忆阻器组成 RC 系统进行 0-9 数字图像的识别任务，通过时分复用原理将数字图像转换成电压脉冲序列，忆阻器将其作为输入信号非线性的映射到高维特征空间，对输入的电压脉冲进行时空特征提取，用忆阻器的瞬态电流响应代表提取到的特征。因此该系统不需要对储层进行训练，只需要训练输出层，利用回声状态网络构建储层状态到输出分类结果的训练模型，训练样本为 200 个，测试样本为 100 个，RC 系统能够正确识别机写 0-9 数字图像信息。实验结果表明，器件可以很好地分离 10 种不同的电压脉冲序列且动态响应稳定。对系统储层状态的差异显著性也进行了统计分析。

# 第 5 章 总结与展望

## 5.1 工作总结

本文曾在第一章提到了忆阻器面临着制造工艺难度大、稳定性不足、计算和控制难度高和难以规模应用等挑战。本文用详实的数据回应了以上挑战，具体为：

首先介绍了基于忆阻器进行神经形态计算领域研究的动态计算，其具有效率高、功耗低等优点。详细介绍了易失性忆阻器相关的阻变机理和重点研究领域，并指出了利用器件本身固有的动态物理过程，将易失性忆阻器和储备池计算结合是一个非常有前途的研究方向。

其次，通过光刻和磁控溅射工艺，制备了五个不同尺寸的 LNO 忆阻器。详细介绍了器件的结构设计、制备工艺流程和测试平台的搭建。

接着，对 LNO 忆阻器进行了阻变特性分析。实验结果表明，器件具有易失性、稳定性和非线性特性。对其阻变原理进行了详细的阐述，构建了一种基于导电细丝原理的导电模型。

最后，利用 LNO 忆阻器的短期记忆和非线性特性，将 5 个 LNO 忆阻器并联作为非线性节点应用在 RC 系统中，成功执行了机写 0-9 数字图像的识别任务。展示了 RC 系统的测试结果，并对系统的性能进行了分析。

本论文证明 LNO 是一种新兴材料，可用于制备易失性忆阻器，并且在神经形态计算领域具有潜在的应用价值。

## 5.2 未来研究展望

本论文构建了基于 LNO 易失性忆阻器的储备池计算系统为实现数字识别任务提供了一个可行的方案。未来还有许多工作值得进一步进行研究：

（1）器件的电阻变化是由于内部的离子迁移导致导电细丝的生成和断裂，因此器件存在长期稳定性和使用寿命方面的问题，还需要对 LNO 薄膜的成膜质量进行进一步的研究用以提高器件的可靠性。

（2）还可以与传统 RC 进行相同数字图像识别任务，比较二者的工作效率和功耗，进一步表明利用易失性忆阻器作为储备池计算的非线性节点对效率和功耗方面的优化。

# 参考文献

# 个人简历 在校期间发表论文及研究成果

硕士期间发表过与本学位论文相关的论文及研究成果如下：

[1]. （第一作者） LiNbO$_3$ dynamic memristors for reservoir computing [J].Frontiers in Neuroscience,2023,17573.

WOS JCR Q2；中国科学院文献情报中心期刊分区表升级版，大类二区，TOP 期刊

[2]. （第四作者） Twist-angle-controlled neutral exciton annihilation in WS$_2$ homostructures [J].Nanoscale,2022,14(14): 5537-5544.

WOS JCR Q1；中国科学院文献情报中心期刊分区表升级版，大类二区，TOP 期刊

[3]. 一种铌酸锂易失性忆阻器及其制备方法[P]. 北京, 申请号：2023101861168, 申请日：2023-03-01, 专利类型：发明, 未公开.